\documentclass[aps,pra,reprint,superscriptaddress,longbibliography]{revtex4-1}

\usepackage{graphicx}
\usepackage{printlen}
\usepackage{color}
\usepackage{amsmath,amssymb}
\usepackage{bm}
\usepackage{upgreek}
\usepackage{xspace}
\usepackage[colorlinks,urlcolor=blue,citecolor=blue,linkcolor=blue]{hyperref}
\usepackage[normalem]{ulem}
\graphicspath{{figures/}}
\usepackage{enumerate}

\newcommand{\Rb}{$^{87}$Rb\xspace}

\newcommand{\reffig}[1]{\mbox{Fig.~\ref{#1}}}

\DeclareMathOperator{\sech}{sech}
\newcommand{\upd}{\text{d}}
\newcommand{\ket}[1]{\left| #1 \right \rangle \xspace}
\newcommand{\spinup}{\ensuremath{\left| \uparrow \right\rangle \xspace}}
\newcommand{\spindown}{\ensuremath{\left| \downarrow \right\rangle \xspace}}

\newcommand{\unit}[1]{\,\mathrm{#1}}

\begin{document}

\title{Precise wavefunction engineering with magnetic resonance}

\author{P.~B.~Wigley}
\affiliation{Department of Quantum Science, The Australian National University, ACT 0200 Australia}

\author{L.~M.~Starkey}
\thanks{L. M. Starkey n\'ee Bennie}
\affiliation{School of Physics and Astronomy, Monash University, Victoria 3800, Australia}

\author{S.~S.~Szigeti}
\affiliation{ARC Centre of Excellence for Engineered Quantum Systems, The University of Queensland, QLD 4072, Australia}

\author{M.~Jasperse}
\affiliation{School of Physics and Astronomy, Monash University, Victoria 3800, Australia}

\author{J.~J.~Hope}
\affiliation{Department of Quantum Science, The Australian National University, ACT 0200 Australia}

\author{L.~D.~Turner}
\affiliation{School of Physics and Astronomy, Monash University, Victoria 3800, Australia}

\author{R.~P.~Anderson}
\email[]{russell.anderson@monash.edu}
\affiliation{School of Physics and Astronomy, Monash University, Victoria 3800, Australia}

\date{\today}

\begin{abstract}
Controlling quantum fluids at their fundamental length scale will yield superlative quantum simulators, precision sensors, and spintronic devices.
This scale is typically below the optical diffraction limit, precluding precise wavefunction engineering using optical potentials alone.
We present a protocol to rapidly control the phase and density of a quantum fluid down to the healing length scale using strong time-dependent coupling between internal states of the fluid in a magnetic field gradient.
We demonstrate this protocol by simulating the creation of a single stationary soliton and double soliton states in a Bose-Einstein condensate with control over the individual soliton positions and trajectories, using experimentally feasible parameters.
Such states are yet to be realized experimentally, and are a path towards engineering soliton gases and exotic topological excitations.
\end{abstract}

\maketitle

\section{Introduction}
Precision quantum engineering is essential for quantum simulation and emulation~\cite{goldman_non-abelian_2009, simon_quantum_2011, britton_engineered_2012, hild_far--equilibrium_2014}, topological quantum computing~\cite{nayak_non-abelian_2008, stern_topological_2013}, spintronics~\cite{han_magnetic_2009, wang_domain_2012} and quantum metrology~\cite{berrada_integrated_2013, strobel_fisher_2014, robins_atom_2013, hush_controlling_2013, Bell:2016}.
Engineering an appropriate system Hamiltonian is only part of the challenge.  It is equally critical to prepare the desired initial wavefunction with high fidelity, and highly desirable to be able to apply coherent unitary operations to that wavefunction.
Ideally these controls would engineer spatial features at the smallest length scale of the system, and complete more rapidly than the fastest dynamics of the uncontrolled fluid; in this Letter we show how to achieve this degree of rapid, precision wavefunction engineering.
Typical approaches fail to achieve this fidelity; optical fields cannot be used to engineer wavefunctions on length scales smaller than the optical wavelength $\lambda$~\cite{denschlag_generating_2000,becker_oscillations_2008,carr_dark-soliton_2001,burger_generation_2002}, while adiabatic relaxation is necessarily slower than the fundamental system timescale~\cite{leanhardt_imprinting_2002}.

Our approach uses strong time-dependent coupling between internal states to control the wavefunction rapidly, with spatial resolution provided by a magnetic field gradient and independent of the optical fields.
We call this \textit{magnetic resonance control} (MRC).
Our exemplar quantum fluid is a pseudospin-$1/2$ Bose-Einstein condensate (BEC) comprised of Zeeman states which are coupled using magnetic dipole transitions.
However, MRC is generally applicable to any spatially-extended quantum system with internal states, provided the splitting can be made spatially dependent and the states admit a time-dependent coupling.
Such systems include Fermi gases~\cite{giorgini_theory_2008}, atoms in optical lattices~\cite{mandel_coherent_2003, schrader_neutral_2004, zhang_manipulation_2006}, and $^3$He films~\cite{godfrin_two-dimensional_1995}.

Optical approaches to wavefunction engineering of BECs have included using off-resonant lasers to induce a local phase shift and engineer a soliton~\cite{denschlag_generating_2000,becker_oscillations_2008}, and transferring angular momentum from a Laguerre-Gaussian beam to engineer a vortex~\cite{andersen_quantized_2006}.
The resolution of these techniques has been limited to $2\unit{\upmu m}$ by the optics used~\cite{denschlag_generating_2000,becker_oscillations_2008} and can never imprint structures finer than the optical diffraction limit, even with superlative optics.
Structures engineered using these techniques have been coarser than the healing length ${\xi = 1/\sqrt{8\pi n a}}$, the shortest distance over which density perturbations exist in a stationary state.
The healing length is ${\xi \sim 270\unit{nm}}$ for a peak number density ${n\sim10^{14}\unit{atoms/cm}^3}$ and $s$-wave scattering length ${a = 5.3\unit{nm}}$.
As an alternative to optical wavefunction engineering, inverting a trapping potential adiabatically has produced vortices~\cite{leanhardt_imprinting_2002} and skyrmions~\cite{leanhardt_coreless_2003}.
While these approaches are not inherently diffraction limited, the experimental apparatus limits these schemes to creating only one type of topological defect, and they must be performed slowly on the system timescale.
An optically-induced magnetic resonance technique was used to create the first vortex in a condensate~\cite{matthews_vortices_1999} and an unstable soliton in a condensate~\cite{anderson_watching_2001}, which we discuss in detail later.

The sharpest stable structure supported by a single component condensate is a stationary dark soliton, hereafter a \textit{black soliton}: a $\pi$ phase step across a density zero of width of order $\xi$.
We use the creation of a single black soliton as a stringent test of our MRC protocol, as a black soliton can only be engineered using a protocol that manipulates both the phase and density of the macroscopic wavefunction with healing length resolution.
Phase engineering alone results in the rapid flow of quantum fluid at the site of the phase gradient as the overly sharp soliton relaxes to the stable solution.  This manifests as the emission of supersonic density peaks~\cite{denschlag_generating_2000}, which we avoid by engineering loss at the solition phase singularity to match the targeted state.
Density engineering can create a diverging group of gray solitons through matter wave interference~\cite{scott_formation_1998, weller_experimental_2008, shomroni_evidence_2009, theocharis_multiple_2010}.
First we describe the protocol used to create a single black soliton in the context of a pseudospin-$1/2$ condensate.
Then we simulate engineering a single black soliton in a \Rb condensate, with experimentally relevant parameters.
Finally we show an extension to this protocol that creates multiple solitons with controlled positions and trajectories.

\section{Model Scheme}
\begin{figure}
    \includegraphics[width=\columnwidth]{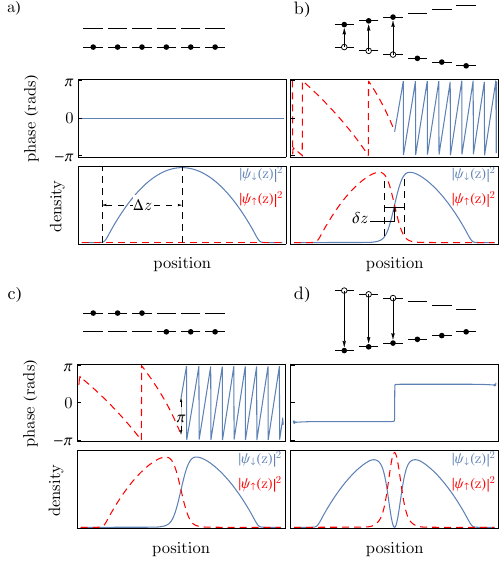}
    \caption{
    \label{fig:mrc_protocol}
        MRC protocol applied to a pseudospin-$1/2$ condensate, showing the position-dependent level splitting, spin state, phases ${\arg\left(\psi_{\downarrow,\uparrow}(z)\right)}$, and densities $\left| \psi_{\downarrow,\uparrow}(z) \right|^2$.
        (a) \textit{Initial wavefunction---}The condensate begins in state $\spindown$ with uniform phase.
        (b) \textit{First pulse---}A magnetic gradient is applied, and a HS pulse transfers the left side of the condensate to state $\spinup$ with slice sharpness $\delta z$.
        The phase of $\spindown$ (blue, solid) acquired during the first pulse is dominated by the gradient, with the shallower phase variation of $\spinup$ (red, dashed) owing to the spin superposition during the pulse.
        (c) \textit{Phase accumulation---}The magnetic gradient is removed, and a uniform $\pi$ phase difference between the two spin components is allowed to accumulate.
        (d) \textit{Second pulse---}The magnetic gradient is inverted and a second pulse transfers the left side back to $\spindown$, imprinting a phase step of $\pi$ onto this component.
        The finite sharpness $\delta z$ of the transferred slice carves a notch in the density $\left| \psi_\downarrow(z) \right|^2$, filled by residual $\spinup$ population.
        MRC enables independent control of the width of the density notch and the height of the phase step.
     }
\end{figure}

Consider a condensate described by a two-component macroscopic wavefunction $\left(\psi_\downarrow(z,t),\psi_\uparrow(z,t)\right)^{\text{T}}$ initially in the spin-down state ($\psi_\uparrow(z,0)=0$) with uniform phase ${\arg\left(\psi_\downarrow(z,0)\right) = 0}$ (\reffig{fig:mrc_protocol}(a)).
The first stage of MRC is to transfer the population of the left side of the condensate to the excited state $\spinup$ (\reffig{fig:mrc_protocol}(b)).
This is achieved with the application of a magnetic field gradient $\upd B / \upd z$ which spatially varies the energy splitting between $\spindown$ and $\spinup$ such that an adiabatic coupling pulse can address a spatial subset of the condensate, as per magnetic resonance imaging~\cite{silver_selective_1985}.
Once the left side of the condensate is in $\spinup$, it accumulates a phase relative to the right side (\reffig{fig:mrc_protocol}(c)).
After the requisite phase shift of $\pi$ has accumulated, the magnetic field gradient is inverted, and a second time-reversed adiabatic pulse is applied, returning the left side of the condensate to the $\spindown$ state (\reffig{fig:mrc_protocol}(d)).
This writes the desired $\pi$ phase step into the wavefunction $\psi_\downarrow(z,t)$.
By inverting the gradient during the second pulse, no net impulse is induced by this Stern-Gerlach force over the duration of the protocol; discussed in detail in section \ref{spin1quasi}.

Our MRC protocol also carves a density notch into $\psi_\downarrow(z,t)$ at the location of the phase step; we can independently control the width of this notch and the accumulated phase step, thereby achieving simultaneous phase and density engineering.
This ${\sim \delta z}$ wide density notch in $\psi_\downarrow(z,t)$ is formed because a finite duration coupling pulse results in a finite edge sharpness $\delta z$ of the transferred slice, leaving some population in $\spinup$ at the edge of the slice after the second coupling pulse.
The sharpness is controlled by choosing the shape of the coupling pulse, discussed in section \ref{adiabaticpulses}.
Rapid density control -- faster than the motional dynamics inherent to the system in the absence of coupling -- can only be achieved by local removal of population.
To create an unfilled soliton, the residual $\spinup$ population is removed using a state-selective transfer to an untrapped state~\cite{ramanathan_partial-transfer_2012}.
Provided that $\delta z \sim \xi$ and the phase step is $\pi$, a single black soliton is created in the scalar condensate.

\section{Adiabatic Pulses\label{adiabaticpulses}}
To address one side of the condensate, we choose a hyperbolic secant coupling pulse~\cite{silver_selective_1985}, which is an adiabatic sweep with a time-dependent Rabi frequency and detuning
\begin{align}
\Omega\left(t\right) &= \Omega_0\sech{\left(\beta\left(t-t_p/2\right)\right)}\\
\Delta\left(t\right) &= \Delta_0\tanh{\left(\beta\left(t-t_p/2\right)\right)}+\Delta_1
\end{align}
defined over the interval of the pulse duration $0 \leq t \leq t_p$ with amplitudes $\Omega_0$ and $\Delta_0$, and sweep rate $\beta$.
The detuning offset $\Delta_1$ sets the center position of the transferred slice.
The magnetic field gradient required to address a slice thickness $\Delta z$ is ${\upd B / \upd z = 2\Delta_0/\left(\gamma \Delta z\right)}$ where $\gamma$ is the gyromagnetic ratio. In order to effectively engineer a black soliton in a particular condensate the pulse parameters must be optimized.

\subsection{Optimal pulse parameters for magnetic resonance control}
Optimizing the pulse parameters is best understood if $\Delta_0$, $\beta$ and $t_p$ are replaced with dimensionless parameters: the normalized pulse bandwidth ${\mu = \Delta_0/\Omega_0}$, adiabaticity ${\Gamma = \Omega_0/\mu\beta}$ and truncation ${\alpha=\sech{\left(\beta t_p/2\right)}}$ (or initial relative amplitude ${\Omega(t=0)/\Omega_0}$ of the pulse).

Together with the description of the MRC protocol time sequence [(1) gradient and pulse; (2) short free evolution with neither gradient nor rf; (3) counter-gradient and counter-pulse], this fully describes the magnetic and rf environment of the quantum fluid during the protocol, although we note that the final `blow away' microwave- or optical-pulse is needed to remove the excited state population and finalize the density engineering. 
However, this removal pulse does not require careful optimization, needing only to be intense enough to remove the excited population in a time short compared to the spin healing time.
In contrast, the parameter optimization for the rf pulse sequence does require carefully constrained optimization: in particular, we seek the lowest peak Rabi frequency $\Omega_0$ and the shallowest magnetic field gradient $dB/dz$ required to produce the desired soliton profile.

The optimization process relies on a relevant measure of success, chosen here to be the resolution of the spatially-dependent transfer, ${R\equiv\Delta z/\delta z}$, where $\Delta z$ is the slice thickness and $\delta z$ is the slice sharpness: taken to be the full width at half maximum (FWHM) and the ${10\%\,\text{--}\,90\%}$ rise distance of the population transfer, respectively.
Engineering a black-soliton in a condensate of healing length $\xi$ and Thomas-Fermi radius $z_\text{TF}$ requires a dimensionless resolution of $R \lesssim 2 z_\text{TF}/5\xi$ \footnote{A black-soliton in an otherwise homogeneous condensate of density $n_0$ has a profile given by $n(z) = n_0 \tanh^2\left(z/\sqrt{2} \xi \right)$, with a corresponding FWHM of $2\sqrt{2} \tanh^{-1}\left(1/\sqrt{2}\right) \xi \simeq 5 \xi /2$. The FWHM of the density modulation resulting from the MRC protcol is $\approx 4\delta z/5$, thus requiring a single-pulse slice sharpness of $\delta z \approx 3\xi$. We set the slice thickness $\Delta z \lesssim 6z_{\text{TF}}/5$ to ensure one side of the pulse is outside the condensate, resulting in the quoted target resolution estimate of $R \lesssim 2 z_\text{TF}/5\xi$}. A resolution greater than this results in sound wave emission from the expanding soliton, while ${R \ll 2 z_{\text{TF}}/5 \xi}$ causes a degeneration into multiple solitons. This ratio is completely determined by the condensate parameters, i.e. trap frequencies, atom number, and scattering length. These parameters also determine the healing time $t_\xi = \sqrt{2} M \xi^2/\hbar$, the time in which a sound wave crosses a healing length. 
This is the shortest time in which the finest features of a condensate evolve due to mean-field interactions. 
If the pulse sequence duration is kept short compared to $t_\xi$, it may be assumed that the order parameter of the condensate behaves like a stationary spin, whereby the internal dynamics (population transfer amongst Zeeman states) dominate the mean-field dynamics and the heuristics below apply. 

\subsubsection{Adiabatic-limited resolution for stationary spins}
The MRC protocol relies on selectively transferring population from one Zeeman state to another.
In the limit of an infinitely long hyperbolic secant pulse applied to stationary spin-1/2 particles, there exists an analytic expression for the fractional population transferred from the spin-down state $\spindown$ to the spin-up state $\spinup$ \cite{silver_selective_1985}:
\begin{equation}
\label{eq:P2_silver}
    P_{\uparrow,\text{final}} = 
    \frac{\rho_{+1}(z)}{\rho(z)} = \frac{\cosh \left(\pi  \Gamma  \mu ^2\right)-\cosh \left(\pi  \Gamma  \mu  \sqrt{\mu ^2-1}\right)}{\cosh \left(\pi  \Gamma  \mu ^2\right)+\cosh \left(\pi  \Gamma  d \mu ^2\right)} \, ,
\end{equation}
where $d=\Delta_z / \Delta_0$ is the normalized detuning offset.
This expression can be used to estimate the resolution $R$ and the pulse fidelity, which we take to be the maximum fractional population transferred to the final Zeeman state after application of the pulse.
To this end, Eq.~(\ref{eq:P2_silver}) can be inverted to find the normalized detuning offset for a given fractional population transferred to the spin-up state,
\begin{equation}
 \begin{aligned}
    d(P_{\uparrow,\text{final}}) = \frac{1}{\pi \Gamma \mu^2} &\cosh^{-1} \left[ \left(P_{\uparrow,\text{final}}^{-1}-1\right) \cosh \left(\pi  \Gamma  \mu ^2\right)\right. \\&\left.- P_{\uparrow,\text{final}}^{-1} \cosh \left(\pi  \Gamma  \mu  \sqrt{\mu ^2-1}\right)\right] \, .
\end{aligned}
\end{equation}
The resolution in the limit of an infinitely long pulse is then
\begin{equation}
\label{eq:R_adiabatic}
    R_{\text{adiabatic}}(\mu, \Gamma) = \frac{\Delta z}{\delta z} = \frac{2 d(0.5)}{d(0.9) - d(0.1)} \, .
\end{equation}
The pulse fidelity is found by setting $d=0$ in Eq.~(\ref{eq:P2_silver}), i.e. the hyperbolic secant pulse transfers population most efficiently when the frequency sweep crosses resonance simultaneously with the maximum coupling amplitude;
\begin{equation}
\begin{aligned}
\label{eq:P1_adiabatic}
    P_{\uparrow,\text{adiabatic}} &\equiv \max_d\left( P_{\uparrow,\text{final}} \right) \\&= 1 - \cosh^2 \left(\tfrac{1}{2} \pi  \Gamma \mu \sqrt{\mu^2-1} \right) \sech^2 \left(\tfrac{1}{2} \pi \Gamma \mu^2 \right) \\&\approx 1 - e^{-\pi \Gamma/2} \, .
\end{aligned}
\end{equation}
The approximation in Eq.~(\ref{eq:P1_adiabatic}) is accurate to within $1\%$ for $\mu, \Gamma > 2$ (and is exact in the limit of $\mu \rightarrow \infty$), and thus $\Gamma$ serves as a good adiabaticity parameter, analogous to that in the Landau-Zener formula.
This population, left behind in the initial state by the finite rate of the sweep, does not participate in the MRC protocol and so reduces the fidelity of the engineered wavefunction by producing an almost uniform background of residual fluid.
It is therefore desirable to minimize this fraction to a specified level, equivalent to choosing a large enough $\Gamma$ for the desired adiabaticity of the pulse.

\subsubsection{Effect of finite-duration pulses}
Eqs.~(\ref{eq:P2_silver})--(\ref{eq:P1_adiabatic}) apply strictly to infinitely long pulses; yet in any experimental scenario the hyperbolic secant pulses are of finite duration, i.e. they are truncated.
Truncating the hyperbolic secant pulse to have a duration $t_p$ results in non-adiabatic, off-resonant Rabi oscillations at the beginning and end of the pulse.
Approximating the coupling at the bounds of the sweep to be that of an unmodulated off-resonant pulse (which is reasonable as both the frequency and amplitude modulation of the hyperbolic secant pulse are slowest here), the off-resonant oscillations in $P_{\downarrow,\text{final}}$ are given by
\begin{align}
\label{eq:P1_asymptotic}
    P_{\downarrow,\text{asymptotic}} &\equiv \frac{\Omega(t=t_p)^2}{\Omega(t=t_p)^2 + \Delta(t=t_p)^2} \\&= \frac{\alpha^2}{\alpha^2 + \mu^2 \left(d + \sqrt{1-\alpha^2} \right)^2} \, .
\end{align}

The residual off-resonant Rabi oscillations manifest as a spatially-varying `roughness' of the engineered wavefunction -- relative to the adiabatic limited shape described in the previous subsection.
To ensure that the resonant pulse fidelity is not limited by this effect, we require $P_{\downarrow,\text{adiabatic}} \leq P_{\downarrow,\text{asymptotic}}$ at $d=0$, resulting in
\begin{equation}
\label{eq:alpha_max_fidelity}
    \alpha < \frac{\mu}{\sqrt{e^{\pi \Gamma/2} +\mu ^2-1}} \, .
\end{equation}
This strictly non-adiabatic coupling does not only perturb the resonant pulse fidelity, but also lowers the resolution of the pulse if $\alpha$ is not chosen to be sufficiently low.
Predicting how small $\alpha$ needs to be so as not to lower the achieved resolution below the adiabatic-limited value in Eq.~(\ref{eq:R_adiabatic}) is non-trivial.
We have found that the resolution limit imparted by finite pulse duration is related to the width of the curve given by Eq.~(\ref{eq:P1_asymptotic}) as a function of normalized detuning $d$ - a Lorentzian with a FWHM of $2\alpha/\mu$.
The resolution with finite duration pulses is $R_{\text{asymptotic}}$, which is sufficiently close to $R_{\text{adiabatic}}$ when
\begin{equation}
\label{eq:alpha_max_resolution}
    \alpha \leq \frac{\mu}{\eta R} \, ,
\end{equation}
where $\eta$ is a design parameter setting the tolerance.
We find that $\eta \approx 30$ typically achieves a pulse truncation limited resolution $R_{\text{asymptotic}}$ within $1\%$ of $R_{\text{adiabatic}}$.
This criterion is \textit{much} stricter than the requirement that $P_{\downarrow,\text{adiabatic}} \leq P_{\downarrow,\text{asymptotic}}$, and thus we use it to choose the pulse truncation $\alpha$.
We now consider how these specifications constrain the optimization of the parameters of the MRC protocol.

\subsubsection{Estimating the optimal adiabatic sweep parameters}
The first step of the MRC protocol is an adiabatic sweep that transfers one spatial half of the condensate to the spin-up state.
Here we assume that the quantum fluid does not move during the protocol; a reasonable approximation provided the pulse sequence is completed faster than the healing time. 

The steps below design an adiabatic sweep with the required resolution $R$ and resonant fidelity $P_{\uparrow,\text{adiabatic}}$.
We keep the design `dimensionless' at first with detunings in units of Rabi frequency, and connect to the experimental parameters later.
\begin{enumerate}
    \item Given a desired resonant fidelity $(1-P_{\downarrow,\text{adiabatic}})$, the adiabaticity parameter is well-approximated by ${\Gamma \approx -(2/\pi) \ln P_{\downarrow,\text{adiabatic}}}$.
    In many experiments a fidelity of 99\% would be considered adequate, and choosing $\Gamma=3$ exceeds this requirement.
    \item The normalized pulse bandwidth $\mu$ is then obtained from $R_{\text{adiabatic}} \approx \sqrt{2} \Gamma \mu^2$, which approximates Eq.~(\ref{eq:R_adiabatic}) to within 1\% for $\Gamma \geq 3$, $\mu \geq 1$.
    This relation implies that a larger $\Gamma$ and correspondingly smaller $\mu$ (or vice versa) would yield the same resolution, and this is correct.
    However, the required effective pulse area $\Omega_0 t_p \propto \Gamma \mu$ becomes larger, so it is better to pick the smallest $\Gamma$ consistent with the desired fidelity and choose a $\mu$ that achieves the resolution $R$ required to write the soliton in question.
    \item For a given design parameter $\eta$, the maximum permissible pulse truncation is determined by Eq.~(\ref{eq:alpha_max_resolution}), $\alpha = 1/(\sqrt{2} \eta \mu \Gamma)$.
\end{enumerate}
This fixes all of the independent dimensionless parameters of the hyperbolic secant pulse. These determine the experimentally relevant quantities as follows:
\begin{enumerate}[i.]
    \item Steps 1--3 above constrain the minimum permissible pulse area
    \begin{equation}
    \label{eq:pulse_area}
        (\Omega_0 t_p)_{\text{min}} = 2 \Gamma \mu \cosh^{-1}\left(\alpha^{-1}\right) .
    \end{equation}
    For a given condensate, we require that $t_p < t_\xi/4$ for the quantum fluid to remain stationary during the pulse sequence, and this sets the required peak Rabi frequency $\Omega_0$.
    \item The pulse bandwidth is then determined via $\Delta_0 = \mu \Omega_0$, with $\mu$ given by Step 2 above.
    \item The sweep rate is then determined via $\beta = \Omega_0^2/(\Delta_0 \Gamma) = \Omega_0 / (\mu \Gamma)$, with $\Gamma$ given by Step 1 above.
    \item The magnetic field gradient must furnish the bandwidth of the sweep across the full spatial extent of the slice, which sets the required gradient via $\gamma \, | dB/dz | \, \Delta z = 2 \Delta_0$.
    Since $\Delta_0 = \mu \Omega_0$ and $\Omega_0$ is given by Eq.~(\ref{eq:pulse_area}), and taking $\Delta z = 6 z_{\text{TF}}/5$, we can alternatively express the required gradient in terms of the healing length alone;
    \begin{equation}
    \label{eq:required_gradient}
        \left|\frac{dB}{dz}\right| = \frac{8 \hbar \cosh^{-1}\left(\alpha^{-1}\right)}{3 \gamma M \xi^3} \, .
    \end{equation}
\end{enumerate}
We note that the heuristic method above serves as a guide to estimate the optimal hyperbolic secant pulse parameters for a given condensate, and illustrates the effect of the dimensionless parameters $\Gamma$, $\mu$, and $\alpha$, their relationship to each other, and how they determine the experimental parameters such as peak Rabi frequency and field gradient.
In practice these parameters are modified slightly upon solving the forward problem: numerical propagation of the Gross-Pitaevskii equation (GPE), discussed in the next section. Indeed experimental application may require real time optimization methods to account for variations from the simulated condensate \cite{wigley_fast_2015}. 

\subsubsection{Gradient-induced motion}
The field gradient that furnishes spatially dependent magnetic resonance also accelerates the different spin components during the pulses -- akin to the Stern-Gerlach effect.
In addition to the self-interaction of the quantum fluid, this effect can also modify the engineered wavefunction in a way not encapsulated by the above heuristics.
The spin component with projection $m \hbar$ will experience an acceleration along $z$ of
\begin{equation}
\label{eq:a_SG}
    a_{\text{SG}} = -\frac{\hbar \gamma m}{M} \frac{d B}{d z} \,
\end{equation}
and will move a distance $\delta z_{\text{SG}} = \tfrac{1}{2} a_{\text{SG}} t_p^2$ during each pulse.
For the amplitude of this motion to be kept below one healing length, $|z_{\text{SG}}| < \xi$ (and recalling that $t_p \lesssim t_{\xi}/4$), one requires that the gradient
\begin{equation}
\label{eq:gradient_SG}
    \left| \frac{d B}{d z} \right| < \frac{16 \hbar}{\gamma M \xi^3} \, .
\end{equation}
This upper bound for the gradient is often close to the required gradient in Eq.~(\ref{eq:required_gradient}), further motivating a careful optimization via simulation of the GPE.
We note that by carefully choosing the direction of both the applied gradient and the resonant position of the hyperbolic secant sweep, this effect can actually serve to increase the resolution achieved relative to the estimates above.

Thus, in order to determine optimal parameters for the MRC scheme, choose an adiabaticity $\Gamma$ so that the population left behind in the initial state is negligible; $\Gamma=3$ ensures more than 99\% transfer at the slice center.
A pulse bandwidth $\mu = 2^{-1/4}(R/\Gamma)^{1/2}$ achieves the needed resolution $R$.
A truncation of $\alpha = 1/(40\,\Gamma \mu)$ gives low abruptness at the start and end of the sweep for order 1\% difference in resolution achieved.
The experimental parameters then follow by fixing the pulse duration $t_p < t_\xi /4$, where the healing time ${t_\xi = \xi/c}$ and $c$ is the speed of sound in the condensate..
Finally, optimization in simulation rapidly converges on the small adjustments needed in parameters, and fixes the inter-pulse time. 

\section{Quasi-1D Gross-Pitaevskii equation with magnetic resonance\label{spin1quasi}}

To demonstrate the MRC protocol, we simulate a coupled psuedospin-1/2 BEC in a magnetic field gradient.
The condensate has an elongated geometry, residing in a cylindrically symmetric potential with radial and axial trap frequencies $\omega_r \gg \omega_z$. A soliton in such a condensate will remain stable for several trap periods~\cite{muryshev_stability_1999}.
As the salient dynamics occur in the axial direction ($z$, the direction of weakest confinement), we use a quasi-1D GPE to reduce the number of simulated spatial dimensions, applying a Thomas-Fermi ansatz to the density along the radial direction~\cite{zhang_effective_2005}.
This is demonstrably more faithful to experimental reality than a strictly one-dimensional GPE, encapsulating the finite radial extent of the condensate; for example, the latter predicts a parabolic single component ground state, which differs to that observed in an elongated condensate due to the lower density at the axial extent of the cloud.
In the elongated `cigar-shaped' regime under consideration, this method reproduces the results of a full three-dimensional GPE simulation without the computational expense.

A cylindrically symmetric two-component BEC has a spinor order parameter
\begin{equation}
\mathbf{\Psi}(r, z) =
    \begin{pmatrix}
    \psi_{\uparrow}\left(r, z\right)\\
    \psi_{\downarrow}\left(r, z\right)
    \end{pmatrix} \, .
\end{equation}
Applying the ansatz of~\cite{zhang_effective_2005}, the wavefunction of each component is
\begin{align}
    \psi_i(r, z) &= \Phi_{\perp,i}(r, \chi_i(z)) f_i(z) \, .
\end{align}
where $\Phi_\perp(r, \chi_i(z))$ encapsulates the radial dependence of the condensate density for each component. These are taken to be Gaussian profiles, with axially-dependent radial scaling factors $\sigma_{\downarrow}(z)$ and $\sigma_{\uparrow}(z)$;
\begin{equation}
    \Phi_{\perp,i}(r, \sigma_i(z)) = \frac{1}{\sqrt{\pi} a_r \sigma_i(z)} \, \exp\left(-\frac{r^2}{a_r \sigma_i(z)^2}\right) \, ,  
\end{equation}
where $a_r = \sqrt{\hbar/(M \omega_r)}$ is the harmonic oscialltor unit length in the radial direction with trapping frequency $\omega_r$.
The fields are normalized such that $\int | \psi_i(r, z) |^2 \, d V = \int_{-\infty}^{\infty} | f_i(z) |^2 \, d z = N_i$, the number of atoms in each of the $i = \spindown$ or $\spinup$ states.
Thus $n_i (r, z) \equiv | \psi_i(r,z) |^2$ is the three-dimensional atomic density with units of $\unit{atoms/m^3}$ and $\rho_i(z) \equiv | f_i(z) |^2$ is the linear density of each component with units of $\unit{atoms/m}$.
From the fields $f_i(z)$ and axially-dependent scaling factors $\sigma_i(z)$ we can calculate experimentally relevant quantities such as the peak atomic density $n_0$, the average density $\langle n \rangle$, and the column density $\tilde{n}(y, z) = \int_{-\infty}^{\infty} n(r=\sqrt{x^2+y^2},z) \,dx$.
Components of the psuedo-spinor $\mathbf{f} = \left(f_{\downarrow}, f_{\uparrow}\right)^T$ obey the coupled quasi-1D GPEs
\begin{align}
\label{eq:quasi1Dgpe}
    i\hbar \frac{\partial f_{\downarrow}}{\partial t} &= \left( -\frac{\hbar^2}{2M} \frac{\partial^2}{\partial z^2} + V + E_{\perp,\downarrow} + g_{\downarrow \downarrow} \eta_{\downarrow \downarrow} \rho_{\downarrow} + g_{\downarrow \uparrow} \eta_{\downarrow \uparrow} \rho_{\uparrow} \right.\notag\\ &\left.\qquad+ \frac{\Delta}{2} \right) f_{\downarrow} + \frac{\Omega}{2} f_{\uparrow} \, , \\
    i\hbar \frac{\partial f_{\uparrow}}{\partial t} &= \left( -\frac{\hbar^2}{2M} \frac{\partial^2}{\partial z^2} + V + E_{\perp,\uparrow} + g_{\uparrow \uparrow} \eta_{\uparrow \uparrow} \rho_{\uparrow} + g_{\downarrow \uparrow} \eta_{\downarrow \uparrow} \rho_{\downarrow} \right.\notag\\ &\left.\qquad- \frac{\Delta}{2} \right) f_{\uparrow} + \frac{\Omega}{2} f_{\downarrow} \, ,
\end{align}
where $M$ is the atomic mass, $V = M \omega_z^2 z^2/2$ is the spin-independent external potential along the axial direction with angular trapping frequency $\omega_z$, $g_{ij} = 4\pi\hbar^2 a_{ij}/M$ describes collisional interactions (with $a_{ii}$ and $a_{i\neq j}$ the intra- and inter-state $s$-wave scattering lengths, respectively), $E_{\perp,i}(z) = \hbar \omega_r \left( 1 + \sigma_i^4 \right)/(2 \sigma_i^2)$ is the transverse mode energy of each component, and $\eta_{ij}(z)$ are interaction scaling factors (with units of $\unit{m^{-2}}$) given by:
\begin{align}
    \eta_{ij}^{-1}(z) &= \pi a_r^2 \left(\sigma_i(z)^2 + \sigma_j(z)^2\right) \, .
    \end{align}
The $z$-dependent radial scaling factors $\sigma_i(z)$ at any given time are determined by a set of auxiliary simultaneous equations arising from the Lagrangian formulation of the quasi-1D GPE~\cite{zhang_effective_2005}:
\begin{equation}
    \sigma_i^4 = 1 + 2 a_{ii} \rho_i + 8 a_{ij} \rho_j \frac{\sigma_i^4}{\left(\sigma_i^2 + \sigma_j^2\right)^2} \: ; \: i \neq j \, .
\end{equation}

The Hamiltonian for linear coupling between the states $\spindown$ and $\spinup$ is
\begin{equation}
\label{eq:H_c}
    H_c = \frac{\hbar}{2}
    \begin{pmatrix}
    \Delta & \Omega \\
    \Omega & -\Delta
    \end{pmatrix} \, ,
\end{equation}
where $\Omega$ is the Rabi frequency for transitions between the two states.
We work in the frame rotating at the instantaneous radiation frequency $\omega$, for which the detuning $\Delta(z,t) = \omega(t) - \gamma B(z,t)$ has spatial dependence via the magnetic field strength $B(z,t) = B(z=0, t) + B_q(t) z$.
Here, $B_q = dB/dz$ is the magnetic field gradient, and $\gamma = \mu_B \left| g_F \right| / \hbar$ is the gyromagnetic ratio for the hyperfine ground state of $^{87}$Rb.
The detuning can alternatively be expressed as $\Delta(z,t) = \Delta(t) - \gamma B_q(t) z$ where $\Delta(t) = \omega(t) - \gamma B(z=0,t)$ is the detuning at $z=0$.
The modulation concomitant to the hyperbolic secant pulse is encapsulated in $\Delta(t)$, and can be affected by changing either the radiation frequency $\omega$ and/or the bias magnetic field $B(z=0)$.

\section{Numerical Simulation}
To simulate MRC for an experimentally relevant system we solve the GPE above for a condensate consisting of $10^4$ \Rb atoms with two spin states $\spinup$ and $\spindown$. 
The simulated condensate is trapped in a harmonic potential with axial frequency ${f_z=20\unit{Hz}}$ and radial frequency ${f_r=1200\unit{Hz}}$, which can be realized, for example, with the optical dipole potential of a $1064\unit{nm}$ laser beam at a power of $20\unit{mW}$ focused to a $10.5\unit{\upmu m}$ waist~\cite{grimm_optical_1999}.
In this trap, the condensate has ${\xi = 150\unit{nm}}$ and Thomas-Fermi radius ${z_{\text{TF}}=39.6\unit{\upmu m}}$.
A black soliton in this condensate has a FWHM of $375\unit{nm}$.

All numerical simulations utilize XMDS2 \cite{dennis_xmds2_2013} with a 4096 point spatial grid and an adaptive Runge-Kutta timestep (ARK45). The number of spatial points was sufficient to resolve the soliton and ensure the simulation was grid independent.

\begin{figure}
    \includegraphics[width=\columnwidth]{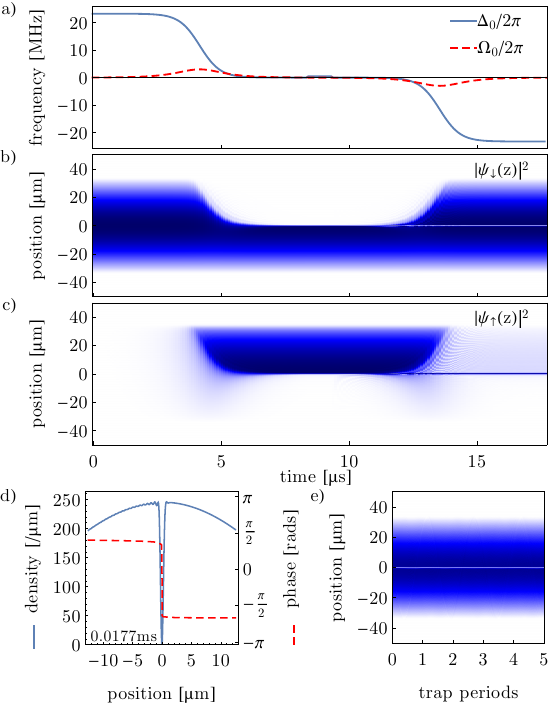}
    \caption{
    \label{fig:single_soliton}
        A single black soliton is engineered using the pulse parameters ${\Omega_0/2\pi = 3.00\unit{MHz}}$, ${\mu=3.87}$, ${\Gamma=3.86}$ and ${\alpha=0.0104}$, resulting in a single pulse duration of ${t_p=8.35\unit{\upmu s}}$ (cf. ${t_{\xi} = 42\unit{\upmu s}}$).
        A gradient of ${\left|\upd B/\upd z\right|=6992.66\unit{G/cm}}$ was applied during the two pulses.
        (a)  The coupling and 
        (b--c) the densities $\left| \psi_{\downarrow,\uparrow}(z) \right|^2$, respectively, during the protocol.
        (d) The density and phase of the condensate at the conclusion of the protocol.
        A $\pi$ phase discontinuity was formed by applying a detuning of ${\Delta_{\phi}/2\pi = 490.0\unit{kHz}}$ during the wait time of ${t_{\phi}=1.0\unit{\upmu s}}$ between pulses.
        The notch in density $\left| \psi \right|^2$ has a FWHM of ${\sim550}\unit{nm}$.
        (e) The total density of the condensate over five trap periods, demonstrating stationary evolution of the black soliton.
        We note that the first hyperbolic secant pulse forms a spin domain wall~\cite{isoshima_spin-domain_1999, watabe_excitation_2012}.
     }
\end{figure}

A single black soliton was engineered in the condensate by choosing the pulse parameters detailed in \reffig{fig:single_soliton}.
Each pulse addressed a slice of thickness ${\Delta z = 1.2z_{TF}}$ and had a slice sharpness of ${\delta z = 674.0\unit{nm}}$.
The resolution of each pulse was thus ${R = 70.4}$, sufficient to produce a single black soliton (cf. ${2z_{\text{TF}}/5 \xi = 105.6}$).
Each pulse required a detuning offset of ${\left|\Delta_1\right| = \mu \Omega_0}$ in order to position the right edge of the slice at ${z=0}$; without this offset, a soliton would be formed at each side of the coupled slice.
The slice thickness ${\Delta z \geq z_{\text{TF}}}$ ensures that the outer edge of the slice is beyond the extent of the condensate.

If no relative phase evolution occurred during the pulses (i.e. in the limit of a very short pulse) then the time $t_{\phi}$ and detuning $\Delta_{\phi}$ between the pulses to effect a relative phase of $\pi$ between the $\ket{\uparrow\,,\downarrow}$ states would be related by ${\Delta_{\phi} t_{\phi} = \pi}$.
However the non-negligible relative phase acquired during each pulse is corrected for by adjusting $\Delta_\phi t_{\phi}$ to generate the requisite \textit{total} phase discontinuity of $\pi$ immediately after the MRC protocol completes.

As shown in \reffig{fig:single_soliton}(d), the protocol successfully engineered both a $\pi$ phase step and a narrow, dark notch in density $\left| \psi_{\downarrow} \right|^2$, filled by residual population in $\spinup$.
To create a single-component soliton we set the populations in $\spinup$ to zero after the pulse sequence.
We quantify the fidelity by the overlap of the engineered wavefunction with the target wavefunction.
We find a normalized overlap of $0.976$ immediately after the blow-away pulse.
The protocol induces a small slosh of the condensate in the trap, which causes the overlap to oscillate at the trap frequency with a mean value of $0.981$.
However the soliton remains almost stationary over more than two trap periods (\reffig{fig:single_soliton}(e)), demonstrating that the protocol indeed produces a single black soliton.
The soliton is tolerant to variations of the pulse parameters, e.g. a $\pm 1\%$ variation of the magnetic field gradient used in \reffig{fig:single_soliton} increases the soliton width by $1\%$, and decreases the mean overlap over 5 trap periods by $< 2\times10^{-4}$, i.e. the change in the soliton stability over this interval is imperceptible.

Immediately after the pulse sequence low-amplitude sound waves develop on either side of the soliton, which we attribute to small imperfections in the prepared soliton state.
A variation of an hyperbolic secant pulse~\cite{garwood_return_2001}, designed to achieve the same pulse resolution in a shorter time, may further minimize sound generation.
We emphasize that the weak sound emission we observe is a much smaller perturbation of the condensate than the superfluid shockwave that is unavoidably created by optical phase imprinting methods.

The peak Rabi frequency used in our simulations corresponds to an oscillating magnetic field amplitude of $4.3\unit{G}$.
This coupling strength, and the field gradients required, are readily achieved in chip trap experiments where current elements are at most millimetres removed from the quantum fluid~\cite{hansel_boseeinstein_2001,fortagh_magnetic_2007,anderson_magnetically_2013}, but may be demanding for some free-space apparatuses where coils and antennas are more than a centimetre away.

\begin{figure}
    \includegraphics[width=\columnwidth]{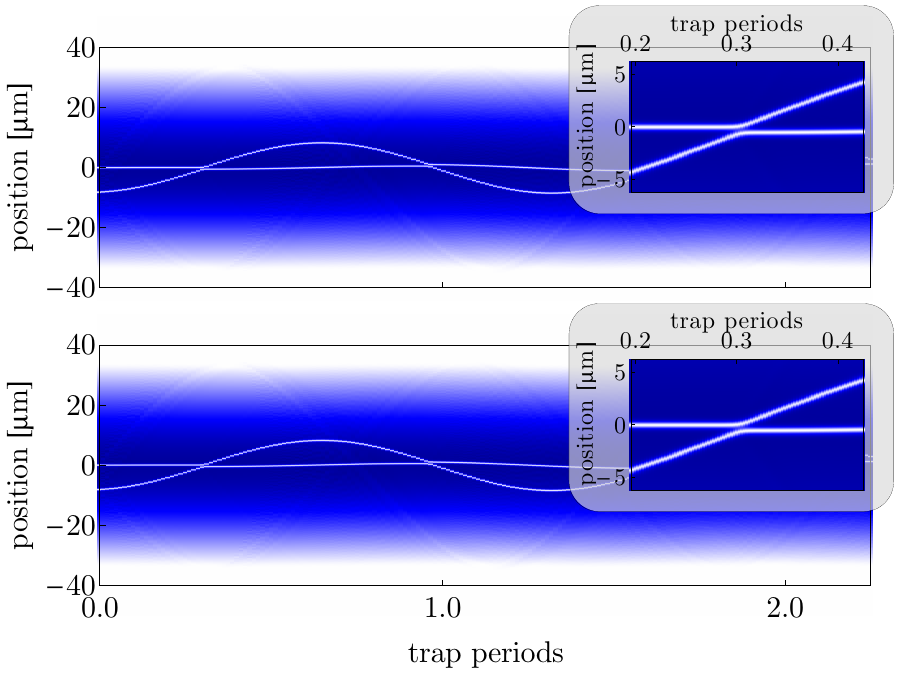}
    \caption{
    \label{fig:double_soliton}
        The MRC protocol can create two solitons by simply decreasing the width of the transferred slice using $\Delta_0$. By varying the offset $\left|\Delta_1\right|$, the position of the solitons can be precisely controlled.
        (a) Two solitons initially located at ${z = -6.6\unit{\upmu m}}$ and ${z = -8.1\unit{\upmu m}}$, formed using ${\mu=0.12}$, ${\Gamma=124.8}$, and an offset of $\left|\Delta_1\right|/(2\pi) = 3.60\unit{MHz}$, exhibit in-phase oscillations.
        (b) If we instead use ${\mu=0.659}$, ${\Gamma=22.69}$ and $\left|\Delta_1\right|/(2\pi) = 1.98\unit{MHz}$, the two solitons are formed at ${z = 0.0\unit{\upmu m}}$ and ${z = -8.1\unit{\upmu m}}$ and oscillate asymmetrically. All other parameters are the same as outlined in \reffig{fig:single_soliton}.
     }
\end{figure}

\section{Extension to multiple solitons}
Finally, we demonstrate the versatility of MRC by engineering double soliton states in the condensate.
Multiple stationary soliton states and their anomalous modes are an analytic continuation of the energy eigenstates of the linear Schr\"odinger equation with harmonic potential, comprised of Hermite-Gauss polynomials~\cite{kivshar_nonlinear_2001,theocharis_multiple_2010}.
Non-stationary multiple dark solitons have been created using phase imprinting methods~\cite{denschlag_generating_2000,stellmer_collisions_2008,becker_inelastic_2013}, and matter wave interference~\cite{weller_experimental_2008}.
We show that MRC provides the precision control needed to create tightly bound or near stationary double solitons, and the extension to more than two solitons is immediate.
The phase of a double soliton forms a top-hat function; such a function could be used as a base unit for constructing more complex atom-optical phase structures such as matter-wave analogues of lenses or mirrors.

As stated previously, our MRC protocol creates a soliton at each edge of the coupled slice.
By simply decreasing the slice thickness, the outer edge is brought within the condensate and two solitons are formed.
To ensure the slice sharpness remains fixed -- preserving the width of the engineered density nodes -- we decrease the slice thickness by decreasing the range of the detuning sweep $\Delta_0$, rather than increasing the magnetic field gradient.
Hence we decrease the normalized pulse bandwidth $\mu$, while proportionally increasing the adiabaticity $\Gamma$; all other parameters remain identical to the single soliton example.
Two solitons are created asymmetrically about the trap center (\reffig{fig:double_soliton}(a)), and periodically exchange all their momentum so that one is stationary while the other oscillates in the trap.
With a different initial separation and mean position, the two solitons can be made to oscillate in phase; this is the lowest anomalous mode of the double soliton state (\reffig{fig:double_soliton}(b)).
Such a double soliton state has never been experimentally realized in a quantum fluid. 

\section{Conclusion}
Magnetic resonance control enables one-dimensional control of both the density and phase of the condensate wavefunction at the healing length and healing time scales.
We have simulated the creation of a single black soliton: a state that has not yet been created in a condensate.
We numerically demonstrated the versatility of MRC, showing it can create double soliton states merely by decreasing the bandwidth swept out by the adiabatic hyperbolic secant pulse.
We anticipate creating a train, or `gas', of solitons using multiple hyperbolic secant pulses, with the same control over their individual trajectories demonstrated above.
Our results establish a path towards magnetic resonance control in higher dimensions, whereby the direction of the magnetic field gradient would be modulated in direct analogy with magnetic resonance imaging, which is fundamentally a one-dimensional technique.
Higher dimensional wavefunction control using MRC could be used to engineer exotic topological excitations such as spin knots in the polar order of a quantum fluid~\cite{kawaguchi_knots_2008}.

\begin{acknowledgments}
This work was supported by the Australian Research Council (ARC) Centre of Excellence for Engineered Quantum Systems (Project No.~CE110001013), the Australian Postgraduate Award Scheme, and ARC grants DP1094399, DP130101613, and FT120100291. S.~S.~S. acknowledges the support of Ian~P.~McCulloch. We are grateful to K.~Helmerson for useful conversations.
\end{acknowledgments}

\bibliography{mrc_soliton}

\begin{thebibliography}{49}%
\makeatletter
\providecommand \@ifxundefined [1]{%
 \@ifx{#1\undefined}
}%
\providecommand \@ifnum [1]{%
 \ifnum #1\expandafter \@firstoftwo
 \else \expandafter \@secondoftwo
 \fi
}%
\providecommand \@ifx [1]{%
 \ifx #1\expandafter \@firstoftwo
 \else \expandafter \@secondoftwo
 \fi
}%
\providecommand \natexlab [1]{#1}%
\providecommand \enquote  [1]{``#1''}%
\providecommand \bibnamefont  [1]{#1}%
\providecommand \bibfnamefont [1]{#1}%
\providecommand \citenamefont [1]{#1}%
\providecommand \href@noop [0]{\@secondoftwo}%
\providecommand \href [0]{\begingroup \@sanitize@url \@href}%
\providecommand \@href[1]{\@@startlink{#1}\@@href}%
\providecommand \@@href[1]{\endgroup#1\@@endlink}%
\providecommand \@sanitize@url [0]{\catcode `\\12\catcode `\$12\catcode
  `\&12\catcode `\#12\catcode `\^12\catcode `\_12\catcode `\%12\relax}%
\providecommand \@@startlink[1]{}%
\providecommand \@@endlink[0]{}%
\providecommand \url  [0]{\begingroup\@sanitize@url \@url }%
\providecommand \@url [1]{\endgroup\@href {#1}{\urlprefix }}%
\providecommand \urlprefix  [0]{URL }%
\providecommand \Eprint [0]{\href }%
\providecommand \doibase [0]{http://dx.doi.org/}%
\providecommand \selectlanguage [0]{\@gobble}%
\providecommand \bibinfo  [0]{\@secondoftwo}%
\providecommand \bibfield  [0]{\@secondoftwo}%
\providecommand \translation [1]{[#1]}%
\providecommand \BibitemOpen [0]{}%
\providecommand \bibitemStop [0]{}%
\providecommand \bibitemNoStop [0]{.\EOS\space}%
\providecommand \EOS [0]{\spacefactor3000\relax}%
\providecommand \BibitemShut  [1]{\csname bibitem#1\endcsname}%
\let\auto@bib@innerbib\@empty
\bibitem [{\citenamefont {Goldman}\ \emph {et~al.}(2009)\citenamefont
  {Goldman}, \citenamefont {Kubasiak}, \citenamefont {Bermudez}, \citenamefont
  {Gaspard}, \citenamefont {Lewenstein},\ and\ \citenamefont
  {Martin-Delgado}}]{goldman_non-abelian_2009}%
  \BibitemOpen
  \bibfield  {author} {\bibinfo {author} {\bibfnamefont {N.}~\bibnamefont
  {Goldman}}, \bibinfo {author} {\bibfnamefont {A.}~\bibnamefont {Kubasiak}},
  \bibinfo {author} {\bibfnamefont {A.}~\bibnamefont {Bermudez}}, \bibinfo
  {author} {\bibfnamefont {P.}~\bibnamefont {Gaspard}}, \bibinfo {author}
  {\bibfnamefont {M.}~\bibnamefont {Lewenstein}}, \ and\ \bibinfo {author}
  {\bibfnamefont {M.~A.}\ \bibnamefont {Martin-Delgado}},\ }\bibfield  {title}
  {\enquote {\bibinfo {title} {Non-Abelian optical lattices: Anomalous quantum
  {Hall} effect and {Dirac} fermions},}\ }\href {\doibase
  10.1103/PhysRevLett.103.035301} {\bibfield  {journal} {\bibinfo  {journal}
  {Phys. Rev. Lett.}\ }\textbf {\bibinfo {volume} {103}},\ \bibinfo {pages}
  {035301} (\bibinfo {year} {2009})}\BibitemShut {NoStop}%
\bibitem [{\citenamefont {Simon}\ \emph {et~al.}(2011)\citenamefont {Simon},
  \citenamefont {Bakr}, \citenamefont {Ma}, \citenamefont {Tai}, \citenamefont
  {Preiss},\ and\ \citenamefont {Greiner}}]{simon_quantum_2011}%
  \BibitemOpen
  \bibfield  {author} {\bibinfo {author} {\bibfnamefont {Jonathan}\
  \bibnamefont {Simon}}, \bibinfo {author} {\bibfnamefont {Waseem~S.}\
  \bibnamefont {Bakr}}, \bibinfo {author} {\bibfnamefont {Ruichao}\
  \bibnamefont {Ma}}, \bibinfo {author} {\bibfnamefont {M.~Eric}\ \bibnamefont
  {Tai}}, \bibinfo {author} {\bibfnamefont {Philipp~M.}\ \bibnamefont
  {Preiss}}, \ and\ \bibinfo {author} {\bibfnamefont {Markus}\ \bibnamefont
  {Greiner}},\ }\bibfield  {title} {\enquote {\bibinfo {title} {Quantum
  simulation of antiferromagnetic spin chains in an optical lattice},}\ }\href
  {\doibase 10.1038/nature09994} {\bibfield  {journal} {\bibinfo  {journal}
  {Nature}\ }\textbf {\bibinfo {volume} {472}},\ \bibinfo {pages} {307--312}
  (\bibinfo {year} {2011})}\BibitemShut {NoStop}%
\bibitem [{\citenamefont {Britton}\ \emph {et~al.}(2012)\citenamefont
  {Britton}, \citenamefont {Sawyer}, \citenamefont {Keith}, \citenamefont
  {Wang}, \citenamefont {Freericks}, \citenamefont {Uys}, \citenamefont
  {Biercuk},\ and\ \citenamefont {Bollinger}}]{britton_engineered_2012}%
  \BibitemOpen
  \bibfield  {author} {\bibinfo {author} {\bibfnamefont {Joseph~W.}\
  \bibnamefont {Britton}}, \bibinfo {author} {\bibfnamefont {Brian~C.}\
  \bibnamefont {Sawyer}}, \bibinfo {author} {\bibfnamefont {Adam~C.}\
  \bibnamefont {Keith}}, \bibinfo {author} {\bibfnamefont {C.-C.~Joseph}\
  \bibnamefont {Wang}}, \bibinfo {author} {\bibfnamefont {James~K.}\
  \bibnamefont {Freericks}}, \bibinfo {author} {\bibfnamefont {Hermann}\
  \bibnamefont {Uys}}, \bibinfo {author} {\bibfnamefont {Michael~J.}\
  \bibnamefont {Biercuk}}, \ and\ \bibinfo {author} {\bibfnamefont {John~J.}\
  \bibnamefont {Bollinger}},\ }\bibfield  {title} {\enquote {\bibinfo {title}
  {Engineered two-dimensional ising interactions in a trapped-ion quantum
  simulator with hundreds of spins},}\ }\href {\doibase 10.1038/nature10981}
  {\bibfield  {journal} {\bibinfo  {journal} {Nature}\ }\textbf {\bibinfo
  {volume} {484}},\ \bibinfo {pages} {489--492} (\bibinfo {year}
  {2012})}\BibitemShut {NoStop}%
\bibitem [{\citenamefont {Hild}\ \emph {et~al.}(2014)\citenamefont {Hild},
  \citenamefont {Fukuhara}, \citenamefont {Schau{\ss}}, \citenamefont {Zeiher},
  \citenamefont {Knap}, \citenamefont {Demler}, \citenamefont {Bloch},\ and\
  \citenamefont {Gross}}]{hild_far--equilibrium_2014}%
  \BibitemOpen
  \bibfield  {author} {\bibinfo {author} {\bibfnamefont {Sebastian}\
  \bibnamefont {Hild}}, \bibinfo {author} {\bibfnamefont {Takeshi}\
  \bibnamefont {Fukuhara}}, \bibinfo {author} {\bibfnamefont {Peter}\
  \bibnamefont {Schau{\ss}}}, \bibinfo {author} {\bibfnamefont {Johannes}\
  \bibnamefont {Zeiher}}, \bibinfo {author} {\bibfnamefont {Michael}\
  \bibnamefont {Knap}}, \bibinfo {author} {\bibfnamefont {Eugene}\ \bibnamefont
  {Demler}}, \bibinfo {author} {\bibfnamefont {Immanuel}\ \bibnamefont
  {Bloch}}, \ and\ \bibinfo {author} {\bibfnamefont {Christian}\ \bibnamefont
  {Gross}},\ }\bibfield  {title} {\enquote {\bibinfo {title}
  {Far-from-equilibrium spin transport in heisenberg quantum magnets},}\ }\href
  {\doibase 10.1103/PhysRevLett.113.147205} {\bibfield  {journal} {\bibinfo
  {journal} {Phys. Rev. Lett.}\ }\textbf {\bibinfo {volume} {113}},\ \bibinfo
  {pages} {147205} (\bibinfo {year} {2014})}\BibitemShut {NoStop}%
\bibitem [{\citenamefont {Nayak}\ \emph {et~al.}(2008)\citenamefont {Nayak},
  \citenamefont {Simon}, \citenamefont {Stern}, \citenamefont {Freedman},\ and\
  \citenamefont {Das~Sarma}}]{nayak_non-abelian_2008}%
  \BibitemOpen
  \bibfield  {author} {\bibinfo {author} {\bibfnamefont {Chetan}\ \bibnamefont
  {Nayak}}, \bibinfo {author} {\bibfnamefont {Steven}\ \bibnamefont {Simon}},
  \bibinfo {author} {\bibfnamefont {Ady}\ \bibnamefont {Stern}}, \bibinfo
  {author} {\bibfnamefont {Michael}\ \bibnamefont {Freedman}}, \ and\ \bibinfo
  {author} {\bibfnamefont {Sankar}\ \bibnamefont {Das~Sarma}},\ }\bibfield
  {title} {\enquote {\bibinfo {title} {Non-Abelian anyons and topological
  quantum computation},}\ }\href {\doibase 10.1103/RevModPhys.80.1083}
  {\bibfield  {journal} {\bibinfo  {journal} {Rev. Mod. Phys.}\ }\textbf
  {\bibinfo {volume} {80}},\ \bibinfo {pages} {1083--1159} (\bibinfo {year}
  {2008})}\BibitemShut {NoStop}%
\bibitem [{\citenamefont {Stern}\ and\ \citenamefont
  {Lindner}(2013)}]{stern_topological_2013}%
  \BibitemOpen
  \bibfield  {author} {\bibinfo {author} {\bibfnamefont {Ady}\ \bibnamefont
  {Stern}}\ and\ \bibinfo {author} {\bibfnamefont {Netanel~H.}\ \bibnamefont
  {Lindner}},\ }\bibfield  {title} {\enquote {\bibinfo {title} {Topological
  quantum computation - from basic concepts to first experiments},}\ }\href
  {\doibase 10.1126/science.1231473} {\bibfield  {journal} {\bibinfo  {journal}
  {Science}\ }\textbf {\bibinfo {volume} {339}},\ \bibinfo {pages} {1179--1184}
  (\bibinfo {year} {2013})}\BibitemShut {NoStop}%
\bibitem [{\citenamefont {Han}\ \emph {et~al.}(2009)\citenamefont {Han},
  \citenamefont {Kim}, \citenamefont {Lee}, \citenamefont {Hermsdoerfer},
  \citenamefont {Schultheiss}, \citenamefont {Leven},\ and\ \citenamefont
  {Hillebrands}}]{han_magnetic_2009}%
  \BibitemOpen
  \bibfield  {author} {\bibinfo {author} {\bibfnamefont {Dong-Soo}\
  \bibnamefont {Han}}, \bibinfo {author} {\bibfnamefont {Sang-Koog}\
  \bibnamefont {Kim}}, \bibinfo {author} {\bibfnamefont {Jun-Young}\
  \bibnamefont {Lee}}, \bibinfo {author} {\bibfnamefont {Sebastian~J.}\
  \bibnamefont {Hermsdoerfer}}, \bibinfo {author} {\bibfnamefont {Helmut}\
  \bibnamefont {Schultheiss}}, \bibinfo {author} {\bibfnamefont {Britta}\
  \bibnamefont {Leven}}, \ and\ \bibinfo {author} {\bibfnamefont {Burkard}\
  \bibnamefont {Hillebrands}},\ }\bibfield  {title} {\enquote {\bibinfo {title}
  {Magnetic domain-wall motion by propagating spin waves},}\ }\href {\doibase
  10.1063/1.3098409} {\bibfield  {journal} {\bibinfo  {journal} {Appl. Phys.
  Lett.}\ }\textbf {\bibinfo {volume} {94}},\ \bibinfo {pages} {112502}
  (\bibinfo {year} {2009})}\BibitemShut {NoStop}%
\bibitem [{\citenamefont {Wang}\ \emph {et~al.}(2012)\citenamefont {Wang},
  \citenamefont {Yan}, \citenamefont {Shen}, \citenamefont {Bauer},\ and\
  \citenamefont {Wang}}]{wang_domain_2012}%
  \BibitemOpen
  \bibfield  {author} {\bibinfo {author} {\bibfnamefont {X.~S.}\ \bibnamefont
  {Wang}}, \bibinfo {author} {\bibfnamefont {P.}~\bibnamefont {Yan}}, \bibinfo
  {author} {\bibfnamefont {Y.~H.}\ \bibnamefont {Shen}}, \bibinfo {author}
  {\bibfnamefont {G.~E.~W.}\ \bibnamefont {Bauer}}, \ and\ \bibinfo {author}
  {\bibfnamefont {X.~R.}\ \bibnamefont {Wang}},\ }\bibfield  {title} {\enquote
  {\bibinfo {title} {Domain wall propagation through spin wave emission},}\
  }\href {\doibase 10.1103/PhysRevLett.109.167209} {\bibfield  {journal}
  {\bibinfo  {journal} {Phys. Rev. Lett.}\ }\textbf {\bibinfo {volume} {109}},\
  \bibinfo {pages} {167209} (\bibinfo {year} {2012})}\BibitemShut {NoStop}%
\bibitem [{\citenamefont {Berrada}\ \emph {et~al.}(2013)\citenamefont
  {Berrada}, \citenamefont {van Frank}, \citenamefont {B{\"u}cker},
  \citenamefont {Schumm}, \citenamefont {Schaff},\ and\ \citenamefont
  {Schmiedmayer}}]{berrada_integrated_2013}%
  \BibitemOpen
  \bibfield  {author} {\bibinfo {author} {\bibfnamefont {T.}~\bibnamefont
  {Berrada}}, \bibinfo {author} {\bibfnamefont {S.}~\bibnamefont {van Frank}},
  \bibinfo {author} {\bibfnamefont {R.}~\bibnamefont {B{\"u}cker}}, \bibinfo
  {author} {\bibfnamefont {T.}~\bibnamefont {Schumm}}, \bibinfo {author}
  {\bibfnamefont {J.-F.}\ \bibnamefont {Schaff}}, \ and\ \bibinfo {author}
  {\bibfnamefont {J.}~\bibnamefont {Schmiedmayer}},\ }\bibfield  {title}
  {\enquote {\bibinfo {title} {Integrated Mach-Zehnder interferometer for
  Bose-Einstein condensates},}\ }\href {http://dx.doi.org/10.1038/ncomms3077}
  {\bibfield  {journal} {\bibinfo  {journal} {Nat. Commun.}\ }\textbf {\bibinfo
  {volume} {4}} (\bibinfo {year} {2013})}\BibitemShut {NoStop}%
\bibitem [{\citenamefont {Strobel}\ \emph {et~al.}(2014)\citenamefont
  {Strobel}, \citenamefont {Muessel}, \citenamefont {Linnemann}, \citenamefont
  {Zibold}, \citenamefont {Hume}, \citenamefont {Pezz{\`e}}, \citenamefont
  {Smerzi},\ and\ \citenamefont {Oberthaler}}]{strobel_fisher_2014}%
  \BibitemOpen
  \bibfield  {author} {\bibinfo {author} {\bibfnamefont {Helmut}\ \bibnamefont
  {Strobel}}, \bibinfo {author} {\bibfnamefont {Wolfgang}\ \bibnamefont
  {Muessel}}, \bibinfo {author} {\bibfnamefont {Daniel}\ \bibnamefont
  {Linnemann}}, \bibinfo {author} {\bibfnamefont {Tilman}\ \bibnamefont
  {Zibold}}, \bibinfo {author} {\bibfnamefont {David~B.}\ \bibnamefont {Hume}},
  \bibinfo {author} {\bibfnamefont {Luca}\ \bibnamefont {Pezz{\`e}}}, \bibinfo
  {author} {\bibfnamefont {Augusto}\ \bibnamefont {Smerzi}}, \ and\ \bibinfo
  {author} {\bibfnamefont {Markus~K.}\ \bibnamefont {Oberthaler}},\ }\bibfield
  {title} {\enquote {\bibinfo {title} {Fisher information and entanglement of
  non-Gaussian spin states},}\ }\href {\doibase 10.1126/science.1250147}
  {\bibfield  {journal} {\bibinfo  {journal} {Science}\ }\textbf {\bibinfo
  {volume} {345}},\ \bibinfo {pages} {424--427} (\bibinfo {year}
  {2014})}\BibitemShut {NoStop}%
\bibitem [{\citenamefont {Robins}\ \emph {et~al.}(2013)\citenamefont {Robins},
  \citenamefont {Altin}, \citenamefont {Debs},\ and\ \citenamefont
  {Close}}]{robins_atom_2013}%
  \BibitemOpen
  \bibfield  {author} {\bibinfo {author} {\bibfnamefont {N.~P.}\ \bibnamefont
  {Robins}}, \bibinfo {author} {\bibfnamefont {P.~A.}\ \bibnamefont {Altin}},
  \bibinfo {author} {\bibfnamefont {J.~E.}\ \bibnamefont {Debs}}, \ and\
  \bibinfo {author} {\bibfnamefont {J.~D.}\ \bibnamefont {Close}},\ }\bibfield
  {title} {\enquote {\bibinfo {title} {Atom lasers: Production, properties and
  prospects for precision inertial measurement},}\ }\href {\doibase
  10.1016/j.physrep.2013.03.006} {\bibfield  {journal} {\bibinfo  {journal}
  {Phys. Rep.}\ }\textbf {\bibinfo {volume} {529}},\ \bibinfo {pages}
  {265--296} (\bibinfo {year} {2013})}\BibitemShut {NoStop}%
\bibitem [{\citenamefont {Hush}\ \emph {et~al.}(2013)\citenamefont {Hush},
  \citenamefont {Szigeti}, \citenamefont {Carvalho},\ and\ \citenamefont
  {Hope}}]{hush_controlling_2013}%
  \BibitemOpen
  \bibfield  {author} {\bibinfo {author} {\bibfnamefont {M.~R.}\ \bibnamefont
  {Hush}}, \bibinfo {author} {\bibfnamefont {S.~S.}\ \bibnamefont {Szigeti}},
  \bibinfo {author} {\bibfnamefont {A.~R.~R.}\ \bibnamefont {Carvalho}}, \ and\
  \bibinfo {author} {\bibfnamefont {J.~J.}\ \bibnamefont {Hope}},\ }\bibfield
  {title} {\enquote {\bibinfo {title} {Controlling spontaneous-emission noise
  in measurement-based feedback cooling of a {B}ose-{E}instein condensate},}\
  }\href {\doibase 10.1088/1367-2630/15/11/113060} {\bibfield  {journal}
  {\bibinfo  {journal} {New J. Phys.}\ }\textbf {\bibinfo {volume} {15}},\
  \bibinfo {pages} {113060} (\bibinfo {year} {2013})}\BibitemShut {NoStop}%
\bibitem [{\citenamefont {Bell}\ \emph {et~al.}(2016)\citenamefont {Bell},
  \citenamefont {Glidden}, \citenamefont {Humbert}, \citenamefont {Bromley},
  \citenamefont {Haine}, \citenamefont {Davis}, \citenamefont {Neely},
  \citenamefont {Baker},\ and\ \citenamefont {Rubinsztein-Dunlop}}]{Bell:2016}%
  \BibitemOpen
  \bibfield  {author} {\bibinfo {author} {\bibfnamefont {Thomas~A}\
  \bibnamefont {Bell}}, \bibinfo {author} {\bibfnamefont {Jake A~P}\
  \bibnamefont {Glidden}}, \bibinfo {author} {\bibfnamefont {Leif}\
  \bibnamefont {Humbert}}, \bibinfo {author} {\bibfnamefont {Michael W~J}\
  \bibnamefont {Bromley}}, \bibinfo {author} {\bibfnamefont {Simon~A}\
  \bibnamefont {Haine}}, \bibinfo {author} {\bibfnamefont {Matthew~J}\
  \bibnamefont {Davis}}, \bibinfo {author} {\bibfnamefont {Tyler~W}\
  \bibnamefont {Neely}}, \bibinfo {author} {\bibfnamefont {Mark~A}\
  \bibnamefont {Baker}}, \ and\ \bibinfo {author} {\bibfnamefont {Halina}\
  \bibnamefont {Rubinsztein-Dunlop}},\ }\bibfield  {title} {\enquote {\bibinfo
  {title} {{B}ose-{E}instein condensation in large time-averaged optical ring
  potentials},}\ }\href {http://stacks.iop.org/1367-2630/18/i=3/a=035003}
  {\bibfield  {journal} {\bibinfo  {journal} {New Journal of Physics}\ }\textbf
  {\bibinfo {volume} {18}},\ \bibinfo {pages} {035003} (\bibinfo {year}
  {2016})}\BibitemShut {NoStop}%
\bibitem [{\citenamefont {Denschlag}\ \emph {et~al.}(2000)\citenamefont
  {Denschlag}, \citenamefont {Simsarian}, \citenamefont {Feder}, \citenamefont
  {Clark}, \citenamefont {Collins}, \citenamefont {Cubizolles}, \citenamefont
  {Deng}, \citenamefont {Hagley}, \citenamefont {Helmerson}, \citenamefont
  {Reinhardt}, \citenamefont {Rolston}, \citenamefont {Schneider},\ and\
  \citenamefont {Phillips}}]{denschlag_generating_2000}%
  \BibitemOpen
  \bibfield  {author} {\bibinfo {author} {\bibfnamefont {J.}~\bibnamefont
  {Denschlag}}, \bibinfo {author} {\bibfnamefont {J.~E.}\ \bibnamefont
  {Simsarian}}, \bibinfo {author} {\bibfnamefont {D.~L.}\ \bibnamefont
  {Feder}}, \bibinfo {author} {\bibfnamefont {Charles~W.}\ \bibnamefont
  {Clark}}, \bibinfo {author} {\bibfnamefont {L.~A.}\ \bibnamefont {Collins}},
  \bibinfo {author} {\bibfnamefont {J.}~\bibnamefont {Cubizolles}}, \bibinfo
  {author} {\bibfnamefont {L.}~\bibnamefont {Deng}}, \bibinfo {author}
  {\bibfnamefont {E.~W.}\ \bibnamefont {Hagley}}, \bibinfo {author}
  {\bibfnamefont {K.}~\bibnamefont {Helmerson}}, \bibinfo {author}
  {\bibfnamefont {W.~P.}\ \bibnamefont {Reinhardt}}, \bibinfo {author}
  {\bibfnamefont {S.~L.}\ \bibnamefont {Rolston}}, \bibinfo {author}
  {\bibfnamefont {B.~I.}\ \bibnamefont {Schneider}}, \ and\ \bibinfo {author}
  {\bibfnamefont {W.~D.}\ \bibnamefont {Phillips}},\ }\bibfield  {title}
  {\enquote {\bibinfo {title} {Generating solitons by phase engineering of a
  {Bose--Einstein} condensate},}\ }\href {\doibase 10.1126/science.287.5450.97}
  {\bibfield  {journal} {\bibinfo  {journal} {Science}\ }\textbf {\bibinfo
  {volume} {287}},\ \bibinfo {pages} {97--101} (\bibinfo {year}
  {2000})}\BibitemShut {NoStop}%
\bibitem [{\citenamefont {Becker}\ \emph {et~al.}(2008)\citenamefont {Becker},
  \citenamefont {Stellmer}, \citenamefont {Soltan-Panahi}, \citenamefont
  {D{\"o}rscher}, \citenamefont {Baumert}, \citenamefont {Richter},
  \citenamefont {Kronj{\"a}ger}, \citenamefont {Bongs},\ and\ \citenamefont
  {Sengstock}}]{becker_oscillations_2008}%
  \BibitemOpen
  \bibfield  {author} {\bibinfo {author} {\bibfnamefont {Christoph}\
  \bibnamefont {Becker}}, \bibinfo {author} {\bibfnamefont {Simon}\
  \bibnamefont {Stellmer}}, \bibinfo {author} {\bibfnamefont {Parvis}\
  \bibnamefont {Soltan-Panahi}}, \bibinfo {author} {\bibfnamefont {S{\"o}ren}\
  \bibnamefont {D{\"o}rscher}}, \bibinfo {author} {\bibfnamefont {Mathis}\
  \bibnamefont {Baumert}}, \bibinfo {author} {\bibfnamefont {Eva-Maria}\
  \bibnamefont {Richter}}, \bibinfo {author} {\bibfnamefont {Jochen}\
  \bibnamefont {Kronj{\"a}ger}}, \bibinfo {author} {\bibfnamefont {Kai}\
  \bibnamefont {Bongs}}, \ and\ \bibinfo {author} {\bibfnamefont {Klaus}\
  \bibnamefont {Sengstock}},\ }\bibfield  {title} {\enquote {\bibinfo {title}
  {Oscillations and interactions of dark and dark-bright solitons in
  Bose-Einstein condensates},}\ }\href {\doibase 10.1038/nphys962} {\bibfield
  {journal} {\bibinfo  {journal} {Nat. Phys.}\ }\textbf {\bibinfo {volume}
  {4}},\ \bibinfo {pages} {496--501} (\bibinfo {year} {2008})}\BibitemShut
  {NoStop}%
\bibitem [{\citenamefont {Carr}\ \emph {et~al.}(2001)\citenamefont {Carr},
  \citenamefont {Brand}, \citenamefont {Burger},\ and\ \citenamefont
  {Sanpera}}]{carr_dark-soliton_2001}%
  \BibitemOpen
  \bibfield  {author} {\bibinfo {author} {\bibfnamefont {L.~D.}\ \bibnamefont
  {Carr}}, \bibinfo {author} {\bibfnamefont {J.}~\bibnamefont {Brand}},
  \bibinfo {author} {\bibfnamefont {S.}~\bibnamefont {Burger}}, \ and\ \bibinfo
  {author} {\bibfnamefont {A.}~\bibnamefont {Sanpera}},\ }\bibfield  {title}
  {\enquote {\bibinfo {title} {Dark-soliton creation in {Bose--Einstein}
  condensates},}\ }\href {\doibase 10.1103/PhysRevA.63.051601} {\bibfield
  {journal} {\bibinfo  {journal} {Phys. Rev. A}\ }\textbf {\bibinfo {volume}
  {63}},\ \bibinfo {pages} {051601} (\bibinfo {year} {2001})}\BibitemShut
  {NoStop}%
\bibitem [{\citenamefont {Burger}\ \emph {et~al.}(2002)\citenamefont {Burger},
  \citenamefont {Carr}, \citenamefont {{\"O}hberg}, \citenamefont {Sengstock},\
  and\ \citenamefont {Sanpera}}]{burger_generation_2002}%
  \BibitemOpen
  \bibfield  {author} {\bibinfo {author} {\bibfnamefont {S.}~\bibnamefont
  {Burger}}, \bibinfo {author} {\bibfnamefont {L.~D.}\ \bibnamefont {Carr}},
  \bibinfo {author} {\bibfnamefont {P.}~\bibnamefont {{\"O}hberg}}, \bibinfo
  {author} {\bibfnamefont {K.}~\bibnamefont {Sengstock}}, \ and\ \bibinfo
  {author} {\bibfnamefont {A.}~\bibnamefont {Sanpera}},\ }\bibfield  {title}
  {\enquote {\bibinfo {title} {Generation and interaction of solitons in
  {Bose--Einstein} condensates},}\ }\href {\doibase 10.1103/PhysRevA.65.043611}
  {\bibfield  {journal} {\bibinfo  {journal} {Phys. Rev. A}\ }\textbf {\bibinfo
  {volume} {65}},\ \bibinfo {pages} {043611} (\bibinfo {year}
  {2002})}\BibitemShut {NoStop}%
\bibitem [{\citenamefont {Leanhardt}\ \emph {et~al.}(2002)\citenamefont
  {Leanhardt}, \citenamefont {G{\"o}rlitz}, \citenamefont {Chikkatur},
  \citenamefont {Kielpinski}, \citenamefont {Shin}, \citenamefont {Pritchard},\
  and\ \citenamefont {Ketterle}}]{leanhardt_imprinting_2002}%
  \BibitemOpen
  \bibfield  {author} {\bibinfo {author} {\bibfnamefont {A.~E.}\ \bibnamefont
  {Leanhardt}}, \bibinfo {author} {\bibfnamefont {A.}~\bibnamefont
  {G{\"o}rlitz}}, \bibinfo {author} {\bibfnamefont {A.~P.}\ \bibnamefont
  {Chikkatur}}, \bibinfo {author} {\bibfnamefont {D.}~\bibnamefont
  {Kielpinski}}, \bibinfo {author} {\bibfnamefont {Y.}~\bibnamefont {Shin}},
  \bibinfo {author} {\bibfnamefont {D.~E.}\ \bibnamefont {Pritchard}}, \ and\
  \bibinfo {author} {\bibfnamefont {W.}~\bibnamefont {Ketterle}},\ }\bibfield
  {title} {\enquote {\bibinfo {title} {Imprinting vortices in a
  {Bose--Einstein} condensate using topological phases},}\ }\href {\doibase
  10.1103/PhysRevLett.89.190403} {\bibfield  {journal} {\bibinfo  {journal}
  {Phys. Rev. Lett.}\ }\textbf {\bibinfo {volume} {89}},\ \bibinfo {pages}
  {190403} (\bibinfo {year} {2002})}\BibitemShut {NoStop}%
\bibitem [{\citenamefont {Giorgini}\ \emph {et~al.}(2008)\citenamefont
  {Giorgini}, \citenamefont {Pitaevskii},\ and\ \citenamefont
  {Stringari}}]{giorgini_theory_2008}%
  \BibitemOpen
  \bibfield  {author} {\bibinfo {author} {\bibfnamefont {Stefano}\ \bibnamefont
  {Giorgini}}, \bibinfo {author} {\bibfnamefont {Lev}\ \bibnamefont
  {Pitaevskii}}, \ and\ \bibinfo {author} {\bibfnamefont {Sandro}\ \bibnamefont
  {Stringari}},\ }\bibfield  {title} {\enquote {\bibinfo {title} {Theory of
  ultracold atomic {Fermi} gases},}\ }\href {\doibase
  10.1103/RevModPhys.80.1215} {\bibfield  {journal} {\bibinfo  {journal} {Rev.
  Mod. Phys.}\ }\textbf {\bibinfo {volume} {80}},\ \bibinfo {pages}
  {1215--1274} (\bibinfo {year} {2008})}\BibitemShut {NoStop}%
\bibitem [{\citenamefont {Mandel}\ \emph {et~al.}(2003)\citenamefont {Mandel},
  \citenamefont {Greiner}, \citenamefont {Widera}, \citenamefont {Rom},
  \citenamefont {H{\"a}nsch},\ and\ \citenamefont
  {Bloch}}]{mandel_coherent_2003}%
  \BibitemOpen
  \bibfield  {author} {\bibinfo {author} {\bibfnamefont {Olaf}\ \bibnamefont
  {Mandel}}, \bibinfo {author} {\bibfnamefont {Markus}\ \bibnamefont
  {Greiner}}, \bibinfo {author} {\bibfnamefont {Artur}\ \bibnamefont {Widera}},
  \bibinfo {author} {\bibfnamefont {Tim}\ \bibnamefont {Rom}}, \bibinfo
  {author} {\bibfnamefont {Theodor~W.}\ \bibnamefont {H{\"a}nsch}}, \ and\
  \bibinfo {author} {\bibfnamefont {Immanuel}\ \bibnamefont {Bloch}},\
  }\bibfield  {title} {\enquote {\bibinfo {title} {Coherent transport of
  neutral atoms in spin-dependent optical lattice potentials},}\ }\href
  {\doibase 10.1103/PhysRevLett.91.010407} {\bibfield  {journal} {\bibinfo
  {journal} {Phys. Rev. Lett.}\ }\textbf {\bibinfo {volume} {91}},\ \bibinfo
  {pages} {010407} (\bibinfo {year} {2003})}\BibitemShut {NoStop}%
\bibitem [{\citenamefont {Schrader}\ \emph {et~al.}(2004)\citenamefont
  {Schrader}, \citenamefont {Dotsenko}, \citenamefont {Khudaverdyan},
  \citenamefont {Miroshnychenko}, \citenamefont {Rauschenbeutel},\ and\
  \citenamefont {Meschede}}]{schrader_neutral_2004}%
  \BibitemOpen
  \bibfield  {author} {\bibinfo {author} {\bibfnamefont {D.}~\bibnamefont
  {Schrader}}, \bibinfo {author} {\bibfnamefont {I.}~\bibnamefont {Dotsenko}},
  \bibinfo {author} {\bibfnamefont {M.}~\bibnamefont {Khudaverdyan}}, \bibinfo
  {author} {\bibfnamefont {Y.}~\bibnamefont {Miroshnychenko}}, \bibinfo
  {author} {\bibfnamefont {A.}~\bibnamefont {Rauschenbeutel}}, \ and\ \bibinfo
  {author} {\bibfnamefont {D.}~\bibnamefont {Meschede}},\ }\bibfield  {title}
  {\enquote {\bibinfo {title} {Neutral atom quantum register},}\ }\href
  {\doibase 10.1103/PhysRevLett.93.150501} {\bibfield  {journal} {\bibinfo
  {journal} {Phys. Rev. Lett.}\ }\textbf {\bibinfo {volume} {93}},\ \bibinfo
  {pages} {150501} (\bibinfo {year} {2004})}\BibitemShut {NoStop}%
\bibitem [{\citenamefont {Zhang}\ \emph {et~al.}(2006)\citenamefont {Zhang},
  \citenamefont {Rolston},\ and\ \citenamefont
  {Das~Sarma}}]{zhang_manipulation_2006}%
  \BibitemOpen
  \bibfield  {author} {\bibinfo {author} {\bibfnamefont {Chuanwei}\
  \bibnamefont {Zhang}}, \bibinfo {author} {\bibfnamefont {S.~L.}\ \bibnamefont
  {Rolston}}, \ and\ \bibinfo {author} {\bibfnamefont {S.}~\bibnamefont
  {Das~Sarma}},\ }\bibfield  {title} {\enquote {\bibinfo {title} {Manipulation
  of single neutral atoms in optical lattices},}\ }\href {\doibase
  10.1103/PhysRevA.74.042316} {\bibfield  {journal} {\bibinfo  {journal} {Phys.
  Rev. A}\ }\textbf {\bibinfo {volume} {74}},\ \bibinfo {pages} {042316}
  (\bibinfo {year} {2006})}\BibitemShut {NoStop}%
\bibitem [{\citenamefont {Godfrin}\ and\ \citenamefont
  {Rapp}(1995)}]{godfrin_two-dimensional_1995}%
  \BibitemOpen
  \bibfield  {author} {\bibinfo {author} {\bibfnamefont {H.}~\bibnamefont
  {Godfrin}}\ and\ \bibinfo {author} {\bibfnamefont {R.E.}\ \bibnamefont
  {Rapp}},\ }\bibfield  {title} {\enquote {\bibinfo {title} {Two-dimensional
  nuclear magnets},}\ }\href {\doibase 10.1080/00018739500101516} {\bibfield
  {journal} {\bibinfo  {journal} {Adv. Phys.}\ }\textbf {\bibinfo {volume}
  {44}},\ \bibinfo {pages} {113--186} (\bibinfo {year} {1995})}\BibitemShut
  {NoStop}%
\bibitem [{\citenamefont {Andersen}\ \emph {et~al.}(2006)\citenamefont
  {Andersen}, \citenamefont {Ryu}, \citenamefont {Clad{\'e}}, \citenamefont
  {Natarajan}, \citenamefont {Vaziri}, \citenamefont {Helmerson},\ and\
  \citenamefont {Phillips}}]{andersen_quantized_2006}%
  \BibitemOpen
  \bibfield  {author} {\bibinfo {author} {\bibfnamefont {M.~F.}\ \bibnamefont
  {Andersen}}, \bibinfo {author} {\bibfnamefont {C.}~\bibnamefont {Ryu}},
  \bibinfo {author} {\bibfnamefont {Pierre}\ \bibnamefont {Clad{\'e}}},
  \bibinfo {author} {\bibfnamefont {Vasant}\ \bibnamefont {Natarajan}},
  \bibinfo {author} {\bibfnamefont {A.}~\bibnamefont {Vaziri}}, \bibinfo
  {author} {\bibfnamefont {K.}~\bibnamefont {Helmerson}}, \ and\ \bibinfo
  {author} {\bibfnamefont {W.~D.}\ \bibnamefont {Phillips}},\ }\bibfield
  {title} {\enquote {\bibinfo {title} {Quantized rotation of atoms from photons
  with orbital angular momentum},}\ }\href {\doibase
  10.1103/PhysRevLett.97.170406} {\bibfield  {journal} {\bibinfo  {journal}
  {Phys. Rev. Lett.}\ }\textbf {\bibinfo {volume} {97}},\ \bibinfo {pages}
  {170406} (\bibinfo {year} {2006})}\BibitemShut {NoStop}%
\bibitem [{\citenamefont {Leanhardt}\ \emph {et~al.}(2003)\citenamefont
  {Leanhardt}, \citenamefont {Shin}, \citenamefont {Kielpinski}, \citenamefont
  {Pritchard},\ and\ \citenamefont {Ketterle}}]{leanhardt_coreless_2003}%
  \BibitemOpen
  \bibfield  {author} {\bibinfo {author} {\bibfnamefont {A.~E.}\ \bibnamefont
  {Leanhardt}}, \bibinfo {author} {\bibfnamefont {Y.}~\bibnamefont {Shin}},
  \bibinfo {author} {\bibfnamefont {D.}~\bibnamefont {Kielpinski}}, \bibinfo
  {author} {\bibfnamefont {D.~E.}\ \bibnamefont {Pritchard}}, \ and\ \bibinfo
  {author} {\bibfnamefont {W.}~\bibnamefont {Ketterle}},\ }\bibfield  {title}
  {\enquote {\bibinfo {title} {Coreless vortex formation in a spinor
  {Bose--Einstein} condensate},}\ }\href {\doibase
  10.1103/PhysRevLett.90.140403} {\bibfield  {journal} {\bibinfo  {journal}
  {Phys. Rev. Lett.}\ }\textbf {\bibinfo {volume} {90}},\ \bibinfo {pages}
  {140403} (\bibinfo {year} {2003})}\BibitemShut {NoStop}%
\bibitem [{\citenamefont {Matthews}\ \emph {et~al.}(1999)\citenamefont
  {Matthews}, \citenamefont {Anderson}, \citenamefont {Haljan}, \citenamefont
  {Hall}, \citenamefont {Wieman},\ and\ \citenamefont
  {Cornell}}]{matthews_vortices_1999}%
  \BibitemOpen
  \bibfield  {author} {\bibinfo {author} {\bibfnamefont {M.~R.}\ \bibnamefont
  {Matthews}}, \bibinfo {author} {\bibfnamefont {B.~P.}\ \bibnamefont
  {Anderson}}, \bibinfo {author} {\bibfnamefont {P.~C.}\ \bibnamefont
  {Haljan}}, \bibinfo {author} {\bibfnamefont {D.~S.}\ \bibnamefont {Hall}},
  \bibinfo {author} {\bibfnamefont {C.~E.}\ \bibnamefont {Wieman}}, \ and\
  \bibinfo {author} {\bibfnamefont {E.~A.}\ \bibnamefont {Cornell}},\
  }\bibfield  {title} {\enquote {\bibinfo {title} {Vortices in a
  {Bose--Einstein} condensate},}\ }\href {\doibase 10.1103/PhysRevLett.83.2498}
  {\bibfield  {journal} {\bibinfo  {journal} {Phys. Rev. Lett.}\ }\textbf
  {\bibinfo {volume} {83}},\ \bibinfo {pages} {2498--2501} (\bibinfo {year}
  {1999})}\BibitemShut {NoStop}%
\bibitem [{\citenamefont {Anderson}\ \emph {et~al.}(2001)\citenamefont
  {Anderson}, \citenamefont {Haljan}, \citenamefont {Regal}, \citenamefont
  {Feder}, \citenamefont {Collins}, \citenamefont {Clark},\ and\ \citenamefont
  {Cornell}}]{anderson_watching_2001}%
  \BibitemOpen
  \bibfield  {author} {\bibinfo {author} {\bibfnamefont {B.~P.}\ \bibnamefont
  {Anderson}}, \bibinfo {author} {\bibfnamefont {P.~C.}\ \bibnamefont
  {Haljan}}, \bibinfo {author} {\bibfnamefont {C.~A.}\ \bibnamefont {Regal}},
  \bibinfo {author} {\bibfnamefont {D.~L.}\ \bibnamefont {Feder}}, \bibinfo
  {author} {\bibfnamefont {L.~A.}\ \bibnamefont {Collins}}, \bibinfo {author}
  {\bibfnamefont {C.~W.}\ \bibnamefont {Clark}}, \ and\ \bibinfo {author}
  {\bibfnamefont {E.~A.}\ \bibnamefont {Cornell}},\ }\bibfield  {title}
  {\enquote {\bibinfo {title} {Watching dark solitons decay into vortex rings
  in a {Bose--Einstein} condensate},}\ }\href {\doibase
  10.1103/PhysRevLett.86.2926} {\bibfield  {journal} {\bibinfo  {journal}
  {Phys. Rev. Lett.}\ }\textbf {\bibinfo {volume} {86}},\ \bibinfo {pages}
  {2926--2929} (\bibinfo {year} {2001})}\BibitemShut {NoStop}%
\bibitem [{\citenamefont {Scott}\ \emph {et~al.}(1998)\citenamefont {Scott},
  \citenamefont {Ballagh},\ and\ \citenamefont
  {Burnett}}]{scott_formation_1998}%
  \BibitemOpen
  \bibfield  {author} {\bibinfo {author} {\bibfnamefont {T.~F.}\ \bibnamefont
  {Scott}}, \bibinfo {author} {\bibfnamefont {R.~J.}\ \bibnamefont {Ballagh}},
  \ and\ \bibinfo {author} {\bibfnamefont {K.}~\bibnamefont {Burnett}},\
  }\bibfield  {title} {\enquote {\bibinfo {title} {Formation of fundamental
  structures in {Bose} - {Einstein} condensates},}\ }\href {\doibase
  10.1088/0953-4075/31/8/001} {\bibfield  {journal} {\bibinfo  {journal} {J.
  Phys. B: At. Mol. Opt. Phys.}\ }\textbf {\bibinfo {volume} {31}},\ \bibinfo
  {pages} {L329} (\bibinfo {year} {1998})}\BibitemShut {NoStop}%
\bibitem [{\citenamefont {Weller}\ \emph {et~al.}(2008)\citenamefont {Weller},
  \citenamefont {Ronzheimer}, \citenamefont {Gross}, \citenamefont {Esteve},
  \citenamefont {Oberthaler}, \citenamefont {Frantzeskakis}, \citenamefont
  {Theocharis},\ and\ \citenamefont {Kevrekidis}}]{weller_experimental_2008}%
  \BibitemOpen
  \bibfield  {author} {\bibinfo {author} {\bibfnamefont {A.}~\bibnamefont
  {Weller}}, \bibinfo {author} {\bibfnamefont {J.~P.}\ \bibnamefont
  {Ronzheimer}}, \bibinfo {author} {\bibfnamefont {C.}~\bibnamefont {Gross}},
  \bibinfo {author} {\bibfnamefont {J.}~\bibnamefont {Esteve}}, \bibinfo
  {author} {\bibfnamefont {M.~K.}\ \bibnamefont {Oberthaler}}, \bibinfo
  {author} {\bibfnamefont {D.~J.}\ \bibnamefont {Frantzeskakis}}, \bibinfo
  {author} {\bibfnamefont {G.}~\bibnamefont {Theocharis}}, \ and\ \bibinfo
  {author} {\bibfnamefont {P.~G.}\ \bibnamefont {Kevrekidis}},\ }\bibfield
  {title} {\enquote {\bibinfo {title} {Experimental observation of oscillating
  and interacting matter wave dark solitons},}\ }\href {\doibase
  10.1103/PhysRevLett.101.130401} {\bibfield  {journal} {\bibinfo  {journal}
  {Phys. Rev. Lett.}\ }\textbf {\bibinfo {volume} {101}},\ \bibinfo {pages}
  {130401} (\bibinfo {year} {2008})}\BibitemShut {NoStop}%
\bibitem [{\citenamefont {Shomroni}\ \emph {et~al.}(2009)\citenamefont
  {Shomroni}, \citenamefont {Lahoud}, \citenamefont {Levy},\ and\ \citenamefont
  {Steinhauer}}]{shomroni_evidence_2009}%
  \BibitemOpen
  \bibfield  {author} {\bibinfo {author} {\bibfnamefont {I.}~\bibnamefont
  {Shomroni}}, \bibinfo {author} {\bibfnamefont {E.}~\bibnamefont {Lahoud}},
  \bibinfo {author} {\bibfnamefont {S.}~\bibnamefont {Levy}}, \ and\ \bibinfo
  {author} {\bibfnamefont {J.}~\bibnamefont {Steinhauer}},\ }\bibfield  {title}
  {\enquote {\bibinfo {title} {Evidence for an oscillating soliton/vortex ring
  by density engineering of a Bose-Einstein condensate},}\ }\href {\doibase
  10.1038/nphys1177} {\bibfield  {journal} {\bibinfo  {journal} {Nat. Phys.}\
  }\textbf {\bibinfo {volume} {5}},\ \bibinfo {pages} {193--197} (\bibinfo
  {year} {2009})}\BibitemShut {NoStop}%
\bibitem [{\citenamefont {Theocharis}\ \emph {et~al.}(2010)\citenamefont
  {Theocharis}, \citenamefont {Weller}, \citenamefont {Ronzheimer},
  \citenamefont {Gross}, \citenamefont {Oberthaler}, \citenamefont
  {Kevrekidis},\ and\ \citenamefont
  {Frantzeskakis}}]{theocharis_multiple_2010}%
  \BibitemOpen
  \bibfield  {author} {\bibinfo {author} {\bibfnamefont {G.}~\bibnamefont
  {Theocharis}}, \bibinfo {author} {\bibfnamefont {A.}~\bibnamefont {Weller}},
  \bibinfo {author} {\bibfnamefont {J.~P.}\ \bibnamefont {Ronzheimer}},
  \bibinfo {author} {\bibfnamefont {C.}~\bibnamefont {Gross}}, \bibinfo
  {author} {\bibfnamefont {M.~K.}\ \bibnamefont {Oberthaler}}, \bibinfo
  {author} {\bibfnamefont {P.~G.}\ \bibnamefont {Kevrekidis}}, \ and\ \bibinfo
  {author} {\bibfnamefont {D.~J.}\ \bibnamefont {Frantzeskakis}},\ }\bibfield
  {title} {\enquote {\bibinfo {title} {Multiple atomic dark solitons in
  cigar-shaped {Bose--Einstein} condensates},}\ }\href {\doibase
  10.1103/PhysRevA.81.063604} {\bibfield  {journal} {\bibinfo  {journal} {Phys.
  Rev. A}\ }\textbf {\bibinfo {volume} {81}},\ \bibinfo {pages} {063604}
  (\bibinfo {year} {2010})}\BibitemShut {NoStop}%
\bibitem [{\citenamefont {Silver}\ \emph {et~al.}(1985)\citenamefont {Silver},
  \citenamefont {Joseph},\ and\ \citenamefont {Hoult}}]{silver_selective_1985}%
  \BibitemOpen
  \bibfield  {author} {\bibinfo {author} {\bibfnamefont {M.~S.}\ \bibnamefont
  {Silver}}, \bibinfo {author} {\bibfnamefont {R.~I.}\ \bibnamefont {Joseph}},
  \ and\ \bibinfo {author} {\bibfnamefont {D.~I.}\ \bibnamefont {Hoult}},\
  }\bibfield  {title} {\enquote {\bibinfo {title} {Selective spin inversion in
  nuclear magnetic resonance and coherent optics through an exact solution of
  the Bloch-Riccati equation},}\ }\href {\doibase 10.1103/PhysRevA.31.2753}
  {\bibfield  {journal} {\bibinfo  {journal} {Phys. Rev. A}\ }\textbf {\bibinfo
  {volume} {31}},\ \bibinfo {pages} {2753--2755} (\bibinfo {year}
  {1985})}\BibitemShut {NoStop}%
\bibitem [{\citenamefont {Ramanathan}\ \emph {et~al.}(2012)\citenamefont
  {Ramanathan}, \citenamefont {Muniz}, \citenamefont {Wright}, \citenamefont
  {Anderson}, \citenamefont {Phillips}, \citenamefont {Helmerson},\ and\
  \citenamefont {Campbell}}]{ramanathan_partial-transfer_2012}%
  \BibitemOpen
  \bibfield  {author} {\bibinfo {author} {\bibfnamefont {Anand}\ \bibnamefont
  {Ramanathan}}, \bibinfo {author} {\bibfnamefont {S{\'e}rgio~R.}\ \bibnamefont
  {Muniz}}, \bibinfo {author} {\bibfnamefont {Kevin~C.}\ \bibnamefont
  {Wright}}, \bibinfo {author} {\bibfnamefont {Russell~P.}\ \bibnamefont
  {Anderson}}, \bibinfo {author} {\bibfnamefont {William~D.}\ \bibnamefont
  {Phillips}}, \bibinfo {author} {\bibfnamefont {Kristian}\ \bibnamefont
  {Helmerson}}, \ and\ \bibinfo {author} {\bibfnamefont {Gretchen~K.}\
  \bibnamefont {Campbell}},\ }\bibfield  {title} {\enquote {\bibinfo {title}
  {Partial-transfer absorption imaging: {A} versatile technique for optimal
  imaging of ultracold gases},}\ }\href {\doibase 10.1063/1.4747163} {\bibfield
   {journal} {\bibinfo  {journal} {Rev. Sci. Instrum.}\ }\textbf {\bibinfo
  {volume} {83}},\ \bibinfo {pages} {083119} (\bibinfo {year}
  {2012})}\BibitemShut {NoStop}%
\bibitem [{Note1()}]{Note1}%
  \BibitemOpen
  \bibinfo {note} {A black-soliton in an otherwise homogeneous condensate of
  density $n_0$ has a profile given by $n(z) = n_0 \protect \qopname \relax
  o{tanh}^2\left (z/\protect \sqrt {2} \xi \right )$, with a corresponding FWHM
  of $2\protect \sqrt {2} \protect \qopname \relax o{tanh}^{-1}\left
  (1/\protect \sqrt {2}\right ) \xi \simeq 5 \xi /2$. The FWHM of the density
  modulation resulting from the MRC protcol is $\approx 4\delta z/5$, thus
  requiring a single-pulse slice sharpness of $\delta z \approx 3\xi $. We set
  the slice thickness $\Delta z \lesssim 6z_{\protect \text {TF}}/5$ to ensure
  one side of the pulse is outside the condensate, resulting in the quoted
  target resolution estimate of $R \lesssim 2 z_\protect \text {TF}/5\xi
  $}\BibitemShut {NoStop}%
\bibitem [{\citenamefont {Wigley}\ \emph {et~al.}(2016)\citenamefont {Wigley},
  \citenamefont {Everitt}, \citenamefont {van~den Hengel}, \citenamefont
  {Bastian}, \citenamefont {Sooriyabandara}, \citenamefont {McDonald},
  \citenamefont {Hardman}, \citenamefont {Quinlivan}, \citenamefont {Manju},
  \citenamefont {Kuhn}, \citenamefont {Petersen}, \citenamefont {Luiten},
  \citenamefont {Hope}, \citenamefont {Robins},\ and\ \citenamefont
  {Hush}}]{wigley_fast_2015}%
  \BibitemOpen
  \bibfield  {author} {\bibinfo {author} {\bibfnamefont {P.~B.}\ \bibnamefont
  {Wigley}}, \bibinfo {author} {\bibfnamefont {P.~J.}\ \bibnamefont {Everitt}},
  \bibinfo {author} {\bibfnamefont {A.}~\bibnamefont {van~den Hengel}},
  \bibinfo {author} {\bibfnamefont {J.~W.}\ \bibnamefont {Bastian}}, \bibinfo
  {author} {\bibfnamefont {M.~A.}\ \bibnamefont {Sooriyabandara}}, \bibinfo
  {author} {\bibfnamefont {G.~D.}\ \bibnamefont {McDonald}}, \bibinfo {author}
  {\bibfnamefont {K.~S.}\ \bibnamefont {Hardman}}, \bibinfo {author}
  {\bibfnamefont {C.~D.}\ \bibnamefont {Quinlivan}}, \bibinfo {author}
  {\bibfnamefont {P.}~\bibnamefont {Manju}}, \bibinfo {author} {\bibfnamefont
  {C.~C.~N.}\ \bibnamefont {Kuhn}}, \bibinfo {author} {\bibfnamefont {I.~R.}\
  \bibnamefont {Petersen}}, \bibinfo {author} {\bibfnamefont {A.~N.}\
  \bibnamefont {Luiten}}, \bibinfo {author} {\bibfnamefont {J.~J.}\
  \bibnamefont {Hope}}, \bibinfo {author} {\bibfnamefont {N.~P.}\ \bibnamefont
  {Robins}}, \ and\ \bibinfo {author} {\bibfnamefont {M.~R.}\ \bibnamefont
  {Hush}},\ }\bibfield  {title} {\enquote {\bibinfo {title} {Fast
  machine-learning online optimization of ultra-cold-atom experiments},}\
  }\href {http://dx.doi.org/10.1038/srep25890} {\bibfield  {journal} {\bibinfo
  {journal} {Scientific Reports}\ }\textbf {\bibinfo {volume} {6}},\ \bibinfo
  {pages} {25890 EP --} (\bibinfo {year} {2016})}\BibitemShut {NoStop}%
\bibitem [{\citenamefont {Muryshev}\ \emph {et~al.}(1999)\citenamefont
  {Muryshev}, \citenamefont {van Linden van~den Heuvell},\ and\ \citenamefont
  {Shlyapnikov}}]{muryshev_stability_1999}%
  \BibitemOpen
  \bibfield  {author} {\bibinfo {author} {\bibfnamefont {A.~E.}\ \bibnamefont
  {Muryshev}}, \bibinfo {author} {\bibfnamefont {H.~B.}\ \bibnamefont {van
  Linden van~den Heuvell}}, \ and\ \bibinfo {author} {\bibfnamefont {G.~V.}\
  \bibnamefont {Shlyapnikov}},\ }\bibfield  {title} {\enquote {\bibinfo {title}
  {Stability of standing matter waves in a trap},}\ }\href {\doibase
  10.1103/PhysRevA.60.R2665} {\bibfield  {journal} {\bibinfo  {journal} {Phys.
  Rev. A}\ }\textbf {\bibinfo {volume} {60}},\ \bibinfo {pages} {R2665--R2668}
  (\bibinfo {year} {1999})}\BibitemShut {NoStop}%
\bibitem [{\citenamefont {Zhang}\ and\ \citenamefont
  {You}(2005)}]{zhang_effective_2005}%
  \BibitemOpen
  \bibfield  {author} {\bibinfo {author} {\bibfnamefont {Wenxian}\ \bibnamefont
  {Zhang}}\ and\ \bibinfo {author} {\bibfnamefont {L.}~\bibnamefont {You}},\
  }\bibfield  {title} {\enquote {\bibinfo {title} {An effective
  quasi-one-dimensional description of a spin-1 atomic condensate},}\ }\href
  {\doibase 10.1103/PhysRevA.71.025603} {\bibfield  {journal} {\bibinfo
  {journal} {Phys. Rev. A}\ }\textbf {\bibinfo {volume} {71}},\ \bibinfo
  {pages} {025603} (\bibinfo {year} {2005})}\BibitemShut {NoStop}%
\bibitem [{\citenamefont {Grimm}\ \emph {et~al.}(1999)\citenamefont {Grimm},
  \citenamefont {Weidemuller},\ and\ \citenamefont
  {Ovchinnikov}}]{grimm_optical_1999}%
  \BibitemOpen
  \bibfield  {author} {\bibinfo {author} {\bibfnamefont {Rudolf}\ \bibnamefont
  {Grimm}}, \bibinfo {author} {\bibfnamefont {Matthias}\ \bibnamefont
  {Weidemuller}}, \ and\ \bibinfo {author} {\bibfnamefont {Yurii~B}\
  \bibnamefont {Ovchinnikov}},\ }\bibfield  {title} {\enquote {\bibinfo {title}
  {Optical dipole traps for neutral atoms},}\ }\href
  {http://arxiv.org/abs/physics/9902072} {\bibfield  {journal} {\bibinfo
  {journal} {Adv. At. Mol. Opt. Phy.}\ ,\ \bibinfo {pages} {95}} (\bibinfo
  {year} {1999})}\BibitemShut {NoStop}%
\bibitem [{\citenamefont {Dennis}\ \emph {et~al.}(2013)\citenamefont {Dennis},
  \citenamefont {Hope},\ and\ \citenamefont {Johnsson}}]{dennis_xmds2_2013}%
  \BibitemOpen
  \bibfield  {author} {\bibinfo {author} {\bibfnamefont {Graham~R.}\
  \bibnamefont {Dennis}}, \bibinfo {author} {\bibfnamefont {Joseph~J.}\
  \bibnamefont {Hope}}, \ and\ \bibinfo {author} {\bibfnamefont {Mattias~T.}\
  \bibnamefont {Johnsson}},\ }\bibfield  {title} {\enquote {\bibinfo {title}
  {{XMDS}2: {Fast}, scalable simulation of coupled stochastic partial
  differential equations},}\ }\href {\doibase 10.1016/j.cpc.2012.08.016}
  {\bibfield  {journal} {\bibinfo  {journal} {Computer Physics Communications}\
  }\textbf {\bibinfo {volume} {184}},\ \bibinfo {pages} {201--208} (\bibinfo
  {year} {2013})}\BibitemShut {NoStop}%
\bibitem [{\citenamefont {Isoshima}\ \emph {et~al.}(1999)\citenamefont
  {Isoshima}, \citenamefont {Machida},\ and\ \citenamefont
  {Ohmi}}]{isoshima_spin-domain_1999}%
  \BibitemOpen
  \bibfield  {author} {\bibinfo {author} {\bibfnamefont {Tomoya}\ \bibnamefont
  {Isoshima}}, \bibinfo {author} {\bibfnamefont {Kazushige}\ \bibnamefont
  {Machida}}, \ and\ \bibinfo {author} {\bibfnamefont {Tetsuo}\ \bibnamefont
  {Ohmi}},\ }\bibfield  {title} {\enquote {\bibinfo {title} {Spin-domain
  formation in spinor {Bose--Einstein} condensation},}\ }\href {\doibase
  10.1103/PhysRevA.60.4857} {\bibfield  {journal} {\bibinfo  {journal} {Phys.
  Rev. A}\ }\textbf {\bibinfo {volume} {60}},\ \bibinfo {pages} {4857--4863}
  (\bibinfo {year} {1999})}\BibitemShut {NoStop}%
\bibitem [{\citenamefont {Watabe}\ \emph {et~al.}(2012)\citenamefont {Watabe},
  \citenamefont {Kato},\ and\ \citenamefont {Ohashi}}]{watabe_excitation_2012}%
  \BibitemOpen
  \bibfield  {author} {\bibinfo {author} {\bibfnamefont {Shohei}\ \bibnamefont
  {Watabe}}, \bibinfo {author} {\bibfnamefont {Yusuke}\ \bibnamefont {Kato}}, \
  and\ \bibinfo {author} {\bibfnamefont {Yoji}\ \bibnamefont {Ohashi}},\
  }\bibfield  {title} {\enquote {\bibinfo {title} {Excitation transport through
  a domain wall in a {Bose--Einstein} condensate},}\ }\href {\doibase
  10.1103/PhysRevA.86.023622} {\bibfield  {journal} {\bibinfo  {journal} {Phys.
  Rev. A}\ }\textbf {\bibinfo {volume} {86}},\ \bibinfo {pages} {023622}
  (\bibinfo {year} {2012})}\BibitemShut {NoStop}%
\bibitem [{\citenamefont {Garwood}\ and\ \citenamefont
  {DelaBarre}(2001)}]{garwood_return_2001}%
  \BibitemOpen
  \bibfield  {author} {\bibinfo {author} {\bibfnamefont {Michael}\ \bibnamefont
  {Garwood}}\ and\ \bibinfo {author} {\bibfnamefont {Lance}\ \bibnamefont
  {DelaBarre}},\ }\bibfield  {title} {\enquote {\bibinfo {title} {The return of
  the frequency sweep: Designing adiabatic pulses for contemporary {NMR}},}\
  }\href {\doibase 10.1006/jmre.2001.2340} {\bibfield  {journal} {\bibinfo
  {journal} {J. Magn. Reson.}\ }\textbf {\bibinfo {volume} {153}},\ \bibinfo
  {pages} {155--177} (\bibinfo {year} {2001})}\BibitemShut {NoStop}%
\bibitem [{\citenamefont {H{\"a}nsel}\ \emph {et~al.}(2001)\citenamefont
  {H{\"a}nsel}, \citenamefont {Hommelhoff}, \citenamefont {H{\"a}nsch},\ and\
  \citenamefont {Reichel}}]{hansel_boseeinstein_2001}%
  \BibitemOpen
  \bibfield  {author} {\bibinfo {author} {\bibfnamefont {W.}~\bibnamefont
  {H{\"a}nsel}}, \bibinfo {author} {\bibfnamefont {P.}~\bibnamefont
  {Hommelhoff}}, \bibinfo {author} {\bibfnamefont {T.~W.}\ \bibnamefont
  {H{\"a}nsch}}, \ and\ \bibinfo {author} {\bibfnamefont {J.}~\bibnamefont
  {Reichel}},\ }\bibfield  {title} {\enquote {\bibinfo {title}
  {{Bose}\textendash{Einstein} condensation on a microelectronic chip},}\
  }\href {\doibase 10.1038/35097032} {\bibfield  {journal} {\bibinfo  {journal}
  {Nature}\ }\textbf {\bibinfo {volume} {413}},\ \bibinfo {pages} {498--501}
  (\bibinfo {year} {2001})}\BibitemShut {NoStop}%
\bibitem [{\citenamefont {Fort\'agh}\ and\ \citenamefont
  {Zimmermann}(2007)}]{fortagh_magnetic_2007}%
  \BibitemOpen
  \bibfield  {author} {\bibinfo {author} {\bibfnamefont {J\'ozsef}\
  \bibnamefont {Fort\'agh}}\ and\ \bibinfo {author} {\bibfnamefont {Claus}\
  \bibnamefont {Zimmermann}},\ }\bibfield  {title} {\enquote {\bibinfo {title}
  {Magnetic microtraps for ultracold atoms},}\ }\href {\doibase
  10.1103/RevModPhys.79.235} {\bibfield  {journal} {\bibinfo  {journal} {Rev.
  Mod. Phys.}\ }\textbf {\bibinfo {volume} {79}},\ \bibinfo {pages} {235--289}
  (\bibinfo {year} {2007})}\BibitemShut {NoStop}%
\bibitem [{\citenamefont {Anderson}\ \emph {et~al.}(2013)\citenamefont
  {Anderson}, \citenamefont {Spielman},\ and\ \citenamefont
  {Juzeli{\=u}nas}}]{anderson_magnetically_2013}%
  \BibitemOpen
  \bibfield  {author} {\bibinfo {author} {\bibfnamefont {Brandon~M.}\
  \bibnamefont {Anderson}}, \bibinfo {author} {\bibfnamefont {I.~B.}\
  \bibnamefont {Spielman}}, \ and\ \bibinfo {author} {\bibfnamefont
  {Gediminas}\ \bibnamefont {Juzeli{\=u}nas}},\ }\bibfield  {title} {\enquote
  {\bibinfo {title} {Magnetically {Generated} {Spin}-{Orbit} {Coupling} for
  {Ultracold} {Atoms}},}\ }\href {\doibase 10.1103/PhysRevLett.111.125301}
  {\bibfield  {journal} {\bibinfo  {journal} {Physical Review Letters}\
  }\textbf {\bibinfo {volume} {111}},\ \bibinfo {pages} {125301} (\bibinfo
  {year} {2013})}\BibitemShut {NoStop}%
\bibitem [{\citenamefont {Kivshar}\ \emph {et~al.}(2001)\citenamefont
  {Kivshar}, \citenamefont {Alexander},\ and\ \citenamefont
  {Turitsyn}}]{kivshar_nonlinear_2001}%
  \BibitemOpen
  \bibfield  {author} {\bibinfo {author} {\bibfnamefont {Yuri~S.}\ \bibnamefont
  {Kivshar}}, \bibinfo {author} {\bibfnamefont {Tristram~J.}\ \bibnamefont
  {Alexander}}, \ and\ \bibinfo {author} {\bibfnamefont {Sergey~K.}\
  \bibnamefont {Turitsyn}},\ }\bibfield  {title} {\enquote {\bibinfo {title}
  {Nonlinear modes of a macroscopic quantum oscillator},}\ }\href {\doibase
  10.1016/S0375-9601(00)00774-X} {\bibfield  {journal} {\bibinfo  {journal}
  {Phys. Lett. A}\ }\textbf {\bibinfo {volume} {278}},\ \bibinfo {pages}
  {225--230} (\bibinfo {year} {2001})}\BibitemShut {NoStop}%
\bibitem [{\citenamefont {Stellmer}\ \emph {et~al.}(2008)\citenamefont
  {Stellmer}, \citenamefont {Becker}, \citenamefont {Soltan-Panahi},
  \citenamefont {Richter}, \citenamefont {D{\"o}rscher}, \citenamefont
  {Baumert}, \citenamefont {Kronj{\"a}ger}, \citenamefont {Bongs},\ and\
  \citenamefont {Sengstock}}]{stellmer_collisions_2008}%
  \BibitemOpen
  \bibfield  {author} {\bibinfo {author} {\bibfnamefont {S.}~\bibnamefont
  {Stellmer}}, \bibinfo {author} {\bibfnamefont {C.}~\bibnamefont {Becker}},
  \bibinfo {author} {\bibfnamefont {P.}~\bibnamefont {Soltan-Panahi}}, \bibinfo
  {author} {\bibfnamefont {E.-M.}\ \bibnamefont {Richter}}, \bibinfo {author}
  {\bibfnamefont {S.}~\bibnamefont {D{\"o}rscher}}, \bibinfo {author}
  {\bibfnamefont {M.}~\bibnamefont {Baumert}}, \bibinfo {author} {\bibfnamefont
  {J.}~\bibnamefont {Kronj{\"a}ger}}, \bibinfo {author} {\bibfnamefont
  {K.}~\bibnamefont {Bongs}}, \ and\ \bibinfo {author} {\bibfnamefont
  {K.}~\bibnamefont {Sengstock}},\ }\bibfield  {title} {\enquote {\bibinfo
  {title} {Collisions of dark solitons in elongated {Bose--Einstein}
  condensates},}\ }\href {\doibase 10.1103/PhysRevLett.101.120406} {\bibfield
  {journal} {\bibinfo  {journal} {Phys. Rev. Lett.}\ }\textbf {\bibinfo
  {volume} {101}},\ \bibinfo {pages} {120406} (\bibinfo {year}
  {2008})}\BibitemShut {NoStop}%
\bibitem [{\citenamefont {Becker}\ \emph {et~al.}(2013)\citenamefont {Becker},
  \citenamefont {Sengstock}, \citenamefont {Schmelcher}, \citenamefont
  {Kevrekidis},\ and\ \citenamefont
  {Carretero-Gonz{\'a}lez}}]{becker_inelastic_2013}%
  \BibitemOpen
  \bibfield  {author} {\bibinfo {author} {\bibfnamefont {C}~\bibnamefont
  {Becker}}, \bibinfo {author} {\bibfnamefont {K}~\bibnamefont {Sengstock}},
  \bibinfo {author} {\bibfnamefont {P}~\bibnamefont {Schmelcher}}, \bibinfo
  {author} {\bibfnamefont {P~G}\ \bibnamefont {Kevrekidis}}, \ and\ \bibinfo
  {author} {\bibfnamefont {R}~\bibnamefont {Carretero-Gonz{\'a}lez}},\
  }\bibfield  {title} {\enquote {\bibinfo {title} {Inelastic collisions of
  solitary waves in anisotropic Bose--Einstein condensates: sling-shot events
  and expanding collision bubbles},}\ }\href
  {http://stacks.iop.org/1367-2630/15/i=11/a=113028} {\bibfield  {journal}
  {\bibinfo  {journal} {New Journal of Physics}\ }\textbf {\bibinfo {volume}
  {15}},\ \bibinfo {pages} {113028} (\bibinfo {year} {2013})}\BibitemShut
  {NoStop}%
\bibitem [{\citenamefont {Kawaguchi}\ \emph {et~al.}(2008)\citenamefont
  {Kawaguchi}, \citenamefont {Nitta},\ and\ \citenamefont
  {Ueda}}]{kawaguchi_knots_2008}%
  \BibitemOpen
  \bibfield  {author} {\bibinfo {author} {\bibfnamefont {Yuki}\ \bibnamefont
  {Kawaguchi}}, \bibinfo {author} {\bibfnamefont {Muneto}\ \bibnamefont
  {Nitta}}, \ and\ \bibinfo {author} {\bibfnamefont {Masahito}\ \bibnamefont
  {Ueda}},\ }\bibfield  {title} {\enquote {\bibinfo {title} {Knots in a spinor
  {Bose--Einstein} condensate},}\ }\href {\doibase
  10.1103/PhysRevLett.100.180403} {\bibfield  {journal} {\bibinfo  {journal}
  {Phys. Rev. Lett.}\ }\textbf {\bibinfo {volume} {100}},\ \bibinfo {pages}
  {180403} (\bibinfo {year} {2008})}\BibitemShut {NoStop}%
\end{thebibliography}%

\end{document}